\documentclass[11pt,prd,aps]{revtex4}

\usepackage{graphicx}
\usepackage{amsmath}
\usepackage{amssymb}
\usepackage{epstopdf}
\usepackage{subfigure}

\newcommand{\be}{\begin{equation}}
\newcommand{\ee}{\end{equation}}
\newcommand{\bea}{\begin{eqnarray}}
\newcommand{\eea}{\end{eqnarray}}
\DeclareMathAlphabet{\mathbfit}{OML}{cmm}{b}{it}

\def\simle{\mathrel{\rlap{\raise 0.511ex \hbox{$<$}}{\lower 0.511ex \hbox{$\sim$}}}}

\newcommand{\Romatre}{Dip.~di Matematica e Fisica, Universit\`a  Roma Tre and INFN, Sezione di Roma Tre,\\ Via della Vasca Navale 84, I-00146 Rome, Italy}
\newcommand{\Valencia}{IFIC and Universidad de Valencia\\ Avenida Blasco Ibanez 13, I-46010 Valencia, Spain}
\newcommand{\RomatreINFN}{Istituto Nazionale di Fisica Nucleare, Sezione di Roma Tre,\\ Via della Vasca Navale 84, I-00146 Rome, Italy}

\begin{document}


\title{Masses and decay constants of $D_{(s)}^*$ and $B_{(s)}^*$ mesons\\[2mm] with $N_f = 2 + 1 + 1$ twisted mass fermions}

\vspace{1cm}
 
 \author{V.~Lubicz} \affiliation{\Romatre}
 \author{A.~Melis} \affiliation{\Valencia}
 \author{S.~Simula} \affiliation{\RomatreINFN}

\author{\includegraphics[draft=false, scale=0.25]{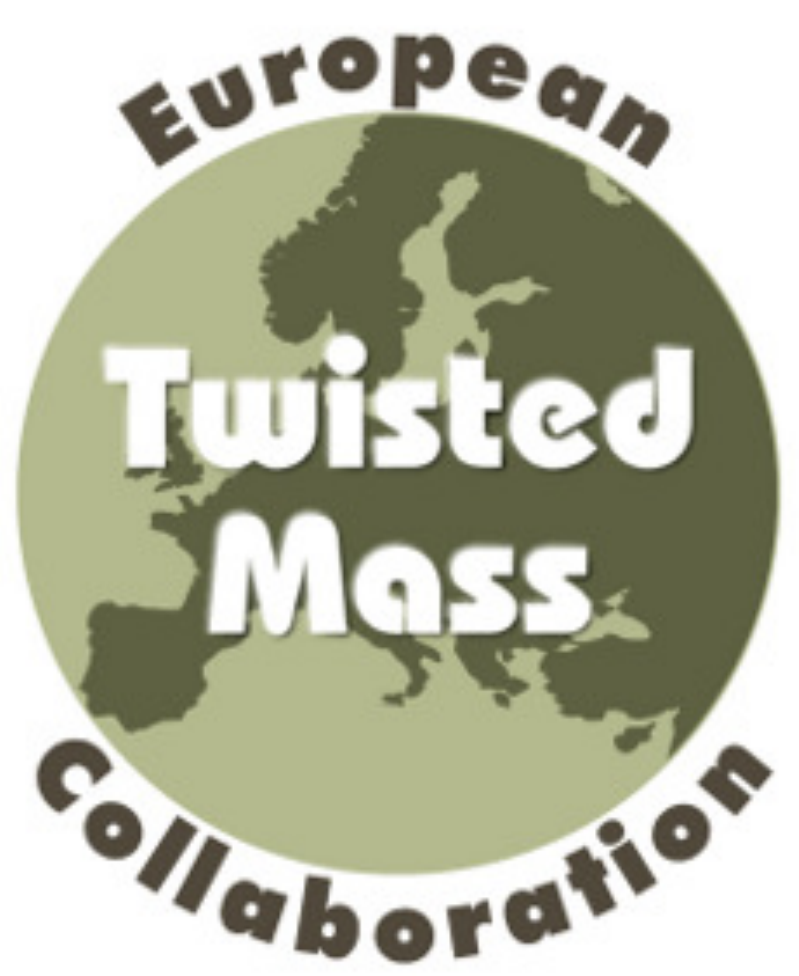}} \noaffiliation

\begin{abstract}
We present a lattice calculation of the masses and decay constants of $D_{(s)}^*$ and $B_{(s)}^*$ mesons using the gauge configurations produced by the European Twisted Mass Collaboration (ETMC)  with $N_f = 2 + 1 + 1$ dynamical quarks at three values of the lattice spacing $a \sim (0.06 - 0.09)$ fm. 
Pion masses are simulated in the range $M_\pi \simeq (210 - 450)$ MeV, while the strange and charm sea-quark masses are close to their physical values. 
We compute the ratios of vector to pseudoscalar masses and decay constants for various values of the heavy-quark mass $m_h$ in the range $0.7 m_c^{\mathrm{phys}} \lesssim m_h \lesssim 3 m_c^{\mathrm{phys}}$. 
In order to reach the physical b-quark mass, we exploit the Heavy Quark Effective Theory prediction that, in the static limit of infinite heavy-quark mass, the considered ratios are equal to one. 
At the physical point our results are: $M_{D^*} / M_{D} = 1.0769(79)$, $M_{D^*_{s}} / M_{D_{s}} = 1.0751(56)$, $ f_{D^*} / f_{D} = 1.078(36),$   $f_{D^*_{s}} / f_{D_{s}} = 1.087(20)$, $M_{B^*} / M_{B} = 1.0078(15)$, $M_{B^*_{s}} / M_{B_{s}} = 1.0083(10)$, $f_{B^*} / f_{B} = 0.958(22)$ and $f_{B^*_{s}} / f_{B_{s}} = 0.974(10)$. 
Combining them with the experimental values of the pseudoscalar meson masses (used as input to fix the quark masses) and the values of the pseudoscalar decay constants calculated by ETMC, we get: $M_{D^*} = 2013(14)~\mathrm{MeV}$, $ M_{D_s^*} = 2116(11)~\mathrm{MeV}$, $f_{D^*} = 223.5(8.4)~\mathrm{MeV}$, $f_{D_s^*} = 268.8(6.6)~\mathrm{MeV}$, $M_{B^*} = 5320.5(7.6)~\mathrm{MeV}$, $M_{B_s^*} = 5411.36(5.3)~\mathrm{MeV}$, $f_{B^*} = 185.9(7.2)~\mathrm{MeV}$ and $f_{B_s^*} = 223.1(5.4)~\mathrm{MeV}$.
\end{abstract}

\maketitle

\newpage

\section{Introduction}
\label{intro}
The decay constants of $D_{(s)}^*$ and $B_{(s)}^*$ mesons play an important role in the phenomenological description of various processes relevant for heavy-flavour physics. For instance, they provide phenomenologically useful descriptions of semileptonic form factors, within the nearest resonance model, and of non-leptonic decay rates in the factorization approximation.

Since the decay modes of $D_{(s)}^*$ and $B_{(s)}^*$ mesons are dominated by the strong and the electromagnetic decays, it is unlikely that their decay constants will be measured directly in the experiments. 
Thus, a non-perturbative approach based on first principles, like lattice QCD simulations, is essential to gain access to these parameters. 
Till now there are only few lattice calculations of the vector-meson decay constants in simulations with either $N_f = 2$ \cite{incriminato,sanfilippo_noi} or $N_f = 2 + 1 (+1)$ \cite{DavisD,DavisB} dynamical quarks. 
Surprisingly a non-negligible difference between present $N_f = 2$ results and those including the strange quark in the sea has been observed \cite{Lubicz:2016bbi}.

The aim of this work is to determine the masses and the decay constants of $D_{(s)}^*$ and $B_{(s)}^*$ mesons using the gauge configurations produced by the European Twisted Mass Collaboration (ETMC)  with $N_f = 2 + 1 + 1$ dynamical quarks.
To this end we will make use of a well known prediction of the Heavy Quark Effective Theory (HQET), namely: in the limit of infinite heavy-quark mass (the static limit) the vector (V) and pseudoscalar (P) heavy-light mesons, which differ only in their internal spin configuration, belong to a doublet of the spin-flavor symmetry and therefore they are degenerate in mass and have the same decay constants.
Consequently, the V to P ratios of masses and decay constants are expected to be equal to one in the static limit, i.e.~$\lim_{m_h \rightarrow \infty} (M_{H^*} / M_H) = 1$ and $\lim_{m_h \rightarrow \infty} (f_{H^*} / f_H) = 1$.

When the heavy quark is either the charm or the beauty, the spin-flavor symmetry is broken and the above ratios deviate from one due to power corrections in $1 / m_h$ and logarithmic radiative corrections.
Since the $b$-quark is still too heavy to be simulated dynamically on present lattices, the HQET asymptotical constraint can be successfully exploited in order to reach the physical $b$-quark sector through an interpolation in the inverse heavy quark mass performed between the static limit and the accessible values of the heavy-quark mass on the lattice.

The paper is organized as follows. In Sec.~\ref{sec:simulations} we summarize the details of the simulations and of the input parameters. In Sec.~\ref{sec:analysis} we start by illustrating the extraction of masses and decay constants calculated on the lattice. Then the results for $D_{(s)}^*$ mesons are presented in Sec.~\ref{sec:FMD} and represent the starting point for the $B_{(s)}^*$-meson analysis to which Sec.~\ref{sec:FMB} is dedicated. Our conclusions are given in Sec.~\ref{sec:concl}.

\section{Simulation details}
\label{sec:simulations}
We used the  gauge ensembles generated by ETMC with $N_f = 2 + 1 + 1$ dynamical quarks \cite{Baron:2010bv,Baron:2010th,Baron:2011sf}. 
In the ETMC setup the gluon interactions are described by the Iwasaki action, while the fermions are regularised with the maximally twisted-mass (Mtm) Wilson lattice formulation.
In order to avoid the mixing of strange and charm quarks in the valence sector we adopted a non-unitary setup in which the valence strange and charm quarks are regularized as Osterwalder-Seiler fermions, while the valence up and down quarks have the same action of the sea.
Working at maximal twist such a setup guarantees an automatic ${\cal{O}}(a)$-improvement and introduces unitarity violations, which however vanish in the continuum limit.

The simulation parameters are summarised in Table~\ref{tab:param}. 
Three values of the lattice spacing are considered, namely $a = 0.0885(36), 0.0815(30)$ and $0.0619(18)$ fm, with the simulated pion mass in the range $M_\pi \simeq (210 - 450)$ MeV. For each lattice spacing different values of the light sea quark mass are studied. 
They are always taken to be equal to the valence up/down degenerate quarks, i.e.~$m^{sea} = m_{u/d}^{val} = m_{u/d}$, while the strange and charm sea-quark masses are close to their physical values \cite{Carrasco:2014cwa}.
The valence quark masses are chosen to be in the ranges: $3 m_{ud}^{\mathrm{phys}} \lesssim m_{u/d} \lesssim 12 m_{u/d}^{\mathrm{phys}}$, $0.7 m_s^{\mathrm{phys}} \lesssim m_s \lesssim 1.2 m_s^{\mathrm{phys}}$ and $0.7 m_c^{\mathrm{phys}} \lesssim m_c \lesssim 1.1 m_c^{\mathrm{phys}}$. 
In order to extrapolate up to the b-quark sector we have also considered higher values of the valence heavy-quark mass in the range $1.1 m_c^{\mathrm{phys}} \lesssim m_h \lesssim 3 m_c^{\mathrm{phys}} \approx 0.7 m_b^{\mathrm{phys}}$. 
The lattice scale has been determined using the experimental value of $f_{\pi^+}$ \cite{Carrasco:2014cwa}, while the physical up/down, strange, charm and bottom quark masses have been fixed in Refs.~\cite{Carrasco:2014cwa,Bussone:2016iua} using the experimental values of $M_\pi$, $M_K$, $M_{D_{(s)}}$ and $M_B$, respectively.

As for the determination of the input parameters, eight branches of the analysis have been implemented in Ref.~\cite{Carrasco:2014cwa}. 
They differ in: 
\begin{itemize}
\item  the strategy for the continuum extrapolation, by choosing as a relative scale parameter either the Sommer parameter $r_0$ or the mass of a fictitious pseudoscalar meson made up of strange(charm)-like quarks;   
\item the chiral extrapolation fit, where the dependence on the light-quark mass is described by either a polynomial expansion or an ansatze based on Chiral Perturbation Theory (ChPT);
\item two methods, denoted as M1 and M2, differing by $\mathcal{O}(a^2)$ effects, for the non-perturbative RI$^\prime$-MOM determination of the renormalization constants (RCs) $Z_m = 1 / Z_P$ and $Z_A$ used in this paper to renormalize the quark masses and the local vector current.  
\end{itemize} 

\begin{table}[tb!] 
\begin{center}
\begin{tabular}{||c|l|c|c|c|c|c||}
\hline
ensemble & $\beta$ & $L^3 \times T$ &$a\mu^{sea}=a\mu_{u/d}$ & $a\mu_s$& $a\mu_c$&  $a\mu_h > a\mu_c$ \\
\hline
\hline
A30.32&$1.90$ & $32^{3}\times 64$  & $0.0030$ & $0.0180$ & $0.21256$ & 0.34583 \\
A40.32& ($a^{-1}\sim$ 2.19 GeV)  &  & $0.0040$ & $0.0220$ & $0.25000$ & 0.40675 \\
A50.32&                                          &  & $0.0050$ & $0.0260$ & $0.29404$ & 0.47840 \\
\cline{1-1}\cline{3-4}
A40.24	& 	& $24^{3}\times 48 $ & $0.0040$ &   		& 			& 0.56267 \\
A60.24	&      &  	                          & $0.0060$ &   		&  			& 0.66178 \\
A80.24	& 	&  	                          & $0.0080$ &    		& 			& 0.77836 \\
A100.24 	&  	&                                & $0.0100$ &    		&  			& 0.91546 \\
\hline
\hline
B25.32& $1.95$ & $32^{3}\times 64$  & $0.0025$ &  $0.0155$ & 0.18705   & 0.30433 \\
B35.32& ($a^{-1}\sim$ 2.50 GeV)    & & $0.0035$ &  $0.0190$ & 0.22000   & 0.35794 \\
B55.32&        				       & & $0.0055$ &  $0.0225$ & 0.25875   & 0.42099 \\
B75.32&        				       & & $0.0075$ &                  &                 & 0.49515 \\
\cline{1-1}\cline{3-4}
B85.24&             & $24^{3}\times 48 $ & $0.0085$ &  	        & 		   & 0.58237 \\
            &            &                                 &                &                  &   		   & 0.68495 \\
            &	          &                                 &                &                  &  		   & 0.80561 \\
\hline
\hline
D15.48& $2.10$ & $48^{3}\times 96$    & $0.0015$ &  $0.0123$ & $0.14454$ & $0.23517$ \\
D20.48&  ($a^{-1}\sim$ 3.23 GeV)     & & $0.0020$ &  $0.0150$ & $0.17000$ & $0.27659$ \\
D30.48&         				         & & $0.0030$ &  $0.0177$ & $0.19995$ & $0.32531$ \\
            &                                             & &                 &                  &  		       & 0.38262 \\
            &                                             & &                 &                  &   		       & 0.45001 \\
            &                                             & &                 &                  &   		       & 0.52928 \\
            &	                                          & &                  &                  &   		       & 0.62252 \\
\hline     					   
\end{tabular}
\end{center}
\caption{\it Simulation parameters for the 15 $ETMC$ gauge ensembles with $N_f = 2 + 1 + 1$ dynamical quarks. For each ensemble we provide the inverse lattice coupling $\beta$, the lattice volumes, the sea and valence bare quark masses. The strange and charm sea-quark masses are close to their physical values.}
\label{tab:param}
\end{table} 

\begin{table}[tb!]
\begin{center}
\scalebox{0.875}{
\begin{tabular}{||c|c|c|c|c|c|c|c|c|c||}
\hline
\multicolumn{1}{||c}{}&\multicolumn{1}{|c|}{$\beta$} & \multicolumn{1}{c|}{ $1^{st}$ }&\multicolumn{1}{c|}{ $2^{nd}$ }&\multicolumn{1}{c|}{ $3^{rd}$ }&\multicolumn{1}{c|}{ $4^{th}$ } & \multicolumn{1}{c|}{ $5^{th}$ }&\multicolumn{1}{c|}{ $6^{th}$ }&\multicolumn{1}{c|}{ $7^{th}$ }&\multicolumn{1}{c||}{$8^{th}$ } \\ \hline  
                         & 1.90 & 2.224(69) &2.191(76) &2.269(87)&2.209(85)& 2.222(67)&2.195(76)   &2.279(90)&2.219(87) \\
$a^{-1}$(GeV)  & 1.95 & 2.416(63) &2.381(73) &2.464(85)&2.400(83)& 2.413(61)&2.384(73)   &2.475(88)&2.411(87) \\
 			& 2.10 & 3.184(59) &3.137(64) &3.248(75)&3.163(75)& 3.181(57)&3.142(65)   &3.262(79)&3.177(78 \\ \cline{1-10}
$m_{u/d}^{\mathrm{phys}}$(GeV) &   &0.00371(13)&0.00386(17)&0.00365(10)&0.00375(13)&0.00362(12)&0.00377(16)&0.00354(9)&0.00363(12) \\ \cline{1-10}
$m_s^{\mathrm{phys}}$(GeV) &   &0.1014(44) &0.1023(39) &0.0992(29)&0.1007(32)& 0.0989(45)&0.0995(39) &0.0962(27)&0.0975(30) \\ \cline{1-10}
$m_c^{\mathrm{phys}}$(GeV) &   & 1.183(34) &1.193(28)  &	1.177(25)&1.219(21)& 1.150(35) &1.1583(27) &1.144(30) &1.181(19) \\ \cline{1-10}
\multicolumn{1}{||c}{$m_b^{\mathrm{phys}}$(GeV) }&\multicolumn{1}{|c|}{}&\multicolumn{4}{c|}{ 5.291(90) }&\multicolumn{4}{c||}{ 5.111(90) } \\ \hline
\multicolumn{1}{||c}{}&\multicolumn{1}{|c|}{1.90}&\multicolumn{4}{c|}{ 0.5290(74)}&\multicolumn{4}{c||}{ 0.5730(42)} \\
\multicolumn{1}{||c}{$Z_P$}&\multicolumn{1}{|c|}{1.95}&\multicolumn{4}{c|}{ 0.5089(34)}&\multicolumn{4}{c||}{ 0.5440(17)} \\
\multicolumn{1}{||c}{}&\multicolumn{1}{|c|}{2.10}&\multicolumn{4}{c|}{ 0.5161(27)}&\multicolumn{4}{c||}{ 0.5420(17)} \\ \hline
\multicolumn{1}{||c}{}&\multicolumn{1}{|c|}{1.90}&\multicolumn{4}{c|}{ 0.7309(86)}&\multicolumn{4}{c||}{ 0.7029(16)} \\
\multicolumn{1}{||c}{$Z_A$}&\multicolumn{1}{|c|}{1.95}&\multicolumn{4}{c|}{ 0.7370(50)}&\multicolumn{4}{c||}{ 0.7139(21)} \\
\multicolumn{1}{||c}{}&\multicolumn{1}{|c|}{2.10}&\multicolumn{4}{c|}{ 0.7621(36)}&\multicolumn{4}{c||}{ 0.7519(21)} \\
\hline
\end{tabular}
}
\end{center}
\caption{\it Input parameters for the eight branches of the analysis of Refs.~\cite{Carrasco:2014cwa,Bussone:2016iua}. The renormalized quark masses and the $RC\,Z_P$ are given in the $\overline{\mathrm{MS}}$ scheme at the renormalization scale of 2 GeV. Branches 1-4 correspond to the use of the RCs determined by the method M1, while branches 5-8 to the RCs obtained with the method M2. With respect to Ref.~\cite{Carrasco:2014cwa} the table includes an update of the values of the lattice spacing and, consequently, of all the other quantities.}
\label{tab:boots}
\end{table} 

The central values and errors of the input parameters for each of the eight branches are evaluated using a bootstrap sampling (of $\mathcal{O}(100)$ events) and are collected in Table~\ref{tab:boots}. 
The eight sets of values represent the input parameters for the present analysis.

\section{Extraction of masses and decay constants}
\label{sec:analysis}
The decay constants of vector and pseudoscalar mesons are defined in terms of the matrix elements of the vector current $\widehat{V}_\mu$ and pseudoscalar density $\widehat{P}$
 \bea
    \label{fV}
    \langle 0 | \widehat{V}_\mu | H^*_\ell(\vec{p}, \lambda) \rangle & = & f_{H_\ell^*} M_{H_\ell^*} \epsilon_\mu^\lambda ~ , \\[2mm]
    \label{fP}
    (m_h + m_\ell) \langle 0 | \widehat{P} | H_\ell(\vec{p}) \rangle & = & p_\mu^{H_\ell} \langle 0 | \widehat{A}_\mu | H_\ell(\vec{p}) \rangle = f_{H_\ell} M_{H_\ell}^2 ~ ,
 \eea
where $M_{H_\ell^{(*)}}$ is the heavy-light meson mass, $m_h$ and $m_\ell$ are the heavy- and light-quark masses with $h = \{c, b\}$ and $\ell = \{u/d, s\}$, $\epsilon_\mu^\lambda$ is the vector meson polarization and $\widehat{A}_\mu$ is the axial current.
In Eq.~(\ref{fP}) the axial Ward-Takahashi identity, which is fulfilled also on the lattice in our Mtm Wilson formulation, has been used.

Ground-state masses and decay constants are determined in lattice QCD by studying two-point correlation functions at large time distances, viz.
 \bea
    \label{CVt}
     C_V(t) & = & \frac{1}{3} \left\langle \sum_{i,\vec{x}} \widehat{V}_i(\vec{x}, t) \widehat{V}_i^{\dagger}(0,0) \right\rangle 
                          \xrightarrow[t \geq t_{\mathrm{min}}]{} \frac{1}{3} \sum_{i, \lambda} \frac{| \langle 0| \widehat{V}_i(0) |H^*_\ell(\lambda) \rangle |^2}{2M_{H_\ell^*}} 
                         \left[ e^{-M_{H_\ell^*} t} + e^{-M_{H_\ell^*} (T - t)} \right] ~ , ~ \\[4mm]
    \label{CPt}
    C_P(t) & = & \left\langle \sum_{\vec{x}} \widehat{P}(\vec{x}, t) \widehat{P}^{\dagger}(0,0) \right\rangle 
                         \xrightarrow[t \geq t_{\mathrm{min}}]{} \frac{| \langle 0| \widehat{P}(0) |H_\ell \rangle |^2} {2M_{H_\ell}} 
                         \left[ e^{-M_{H_\ell} t} + e^{-M_{H_\ell} (T - t)} \right] ~ ,  ~ 
 \eea
where $t_{\mathrm{min}}$ stands for the minimum time-distance at which the ground state can be considered isolated.
In Eqs.~(\ref{CVt}-\ref{CPt}) we employ the local versions of both the vector current $\widehat{V}_i \equiv Z_A \overline{h} \gamma_i \ell$ and the pseudoscalar density $\widehat{P} \equiv Z_P \overline{h} \gamma_5 \ell$, which in our Mtm setup renormalize multiplicatively with the RCs $Z_A$ and $Z_P$, respectively, once opposite values of the Wilson $r$-parameter are adopted for the two valence quarks.
Since at maximal twist the mass RC is given by $Z_m = 1 / Z_P$, the operator $(m_h + m_\ell) \widehat{P}$ becomes $(\mu_h + \mu_\ell) \overline{h} \gamma_5 \ell$, where $\mu_h$ and $\mu_\ell$ are bare quark masses, which is renormalization group invariant and does not require any RC.

The correlation functions (\ref{CVt}) and (\ref{CPt}), based on local interpolating fields, contain both the vector and pseudoscalar ground-state meson masses, $M_{H_\ell}$ and $M_{H^*_\ell}$, as well as the matrix elements required to compute $f_{H_\ell}$ and $f_{H_\ell^*}$ from Eqs.~(\ref{fV}-\ref{fP}). 
In order to improve the determination of the above quantities we have analyzed the whole set of correlation functions given by the combinations of local interpolating operators with those obtained from a Gaussian smearing procedure at both the sink and the source, namely $C_{P,V}^{LL}, C_{P,V}^{LS}, C_{P,V}^{SL}$ and $C_{P,V}^{SS}$, where $L$ and $S$ denote local and smeared operators, respectively.

For the reasons explained in the Introduction we have considered the following ratios 
 \bea
    \label{RM}
    R_{\ell}^M(m_h) & = & \frac{M_{H_\ell^*}}{M_{H_\ell}} ~ , \\[2mm]
    \label{Rf}
    R_{\ell}^f(m_h) & = & \frac{f_{H_\ell^*}}{f_{H_\ell}} ~ ,
 \eea
which go to unity in the static limit. 
Considering these ratios has also the benefit that the uncertainties due to the chiral and continuum extrapolations are significantly reduced with respect to the case of the individual V and P masses/decay constants (see next Sections).

In order to determine the mass ratio (\ref{RM}) we have considered the ratio of the effective masses of V and P correlators, namely
 \be
       \label{Reff}
      \overline{R}_{eff}(t) \equiv \frac{M_{eff}^V(t)}{M_{eff}^P(t)} \xrightarrow[t \geq t_{\mathrm{min}}]{} R_{\ell}^M ~ ,
 \ee
 where
  \be
      \label{Meff}
      M_{eff}^{P, V}(t) \equiv \mbox{arcosh}\left[ \frac{C_{P, V}(t - 1) + C_{P, V}(t + 1)}{2 C_{P, V}(t )} \right] \xrightarrow[t \geq t_{\mathrm{min}}]{} M_{H_\ell^{(*)}} ~ .
   \ee
As for the ratio (\ref{Rf}), in the case of LL correlators we have constructed the quantity
 \be
      \label{Rf_LL}
      \overline{R}_f^{LL}(t) \equiv \frac{\mbox{sinh}(aM_{H_\ell})}{a\mu_h + a\mu_\ell} \sqrt{ \frac{C_V^{LL}(t)}{C_P^{LL}(t)} \frac{M_{H_\ell}}{M_{H_\ell^*}} 
                                                         \frac{e^{-M_{H_\ell} t} + e^{-M_{H_\ell} (T - t)}}{e^{-M_{H_\ell^*} t} + e^{-M_{H_\ell^*} (T - t)}} } 
                                                         \xrightarrow[t \geq t_{\mathrm{min}}]{} R_{\ell}^f ~ ,
 \ee
where the factor $\mbox{sinh}(aM_{H_\ell})$ comes from the temporal derivative of the axial current on the lattice in Eq.~(\ref{fP}).
In the case of smeared correlators we have used the ratio
 \be
      \label{Rf_SL}
      \overline{R}_f^{SL}(t) \equiv \frac{\mbox{sinh}(aM_{H_\ell})}{a\mu_h + a\mu_\ell} \frac{C_V^{SL}(t)}{C_P^{SL}(t)} 
                                                        \sqrt{ \frac{C_P^{SS}(t)}{C_V^{SS}(t)} \frac{M_{H_\ell}}{M_{H_\ell^*}} 
                                                        \frac{e^{-M_{H_\ell} t} + e^{-M_{H_\ell} (T - t)}}{e^{-M_{H_\ell^*} t} + e^{-M_{H_\ell^*} (T - t)}} } 
                                                        \xrightarrow[t \geq t_{\mathrm{min}}]{} R_{\ell}^f ~ .
 \ee

\begin{figure}[tb!]
\centering
\scalebox{1}{\subfigure[\it]{\includegraphics[width=8.15cm]{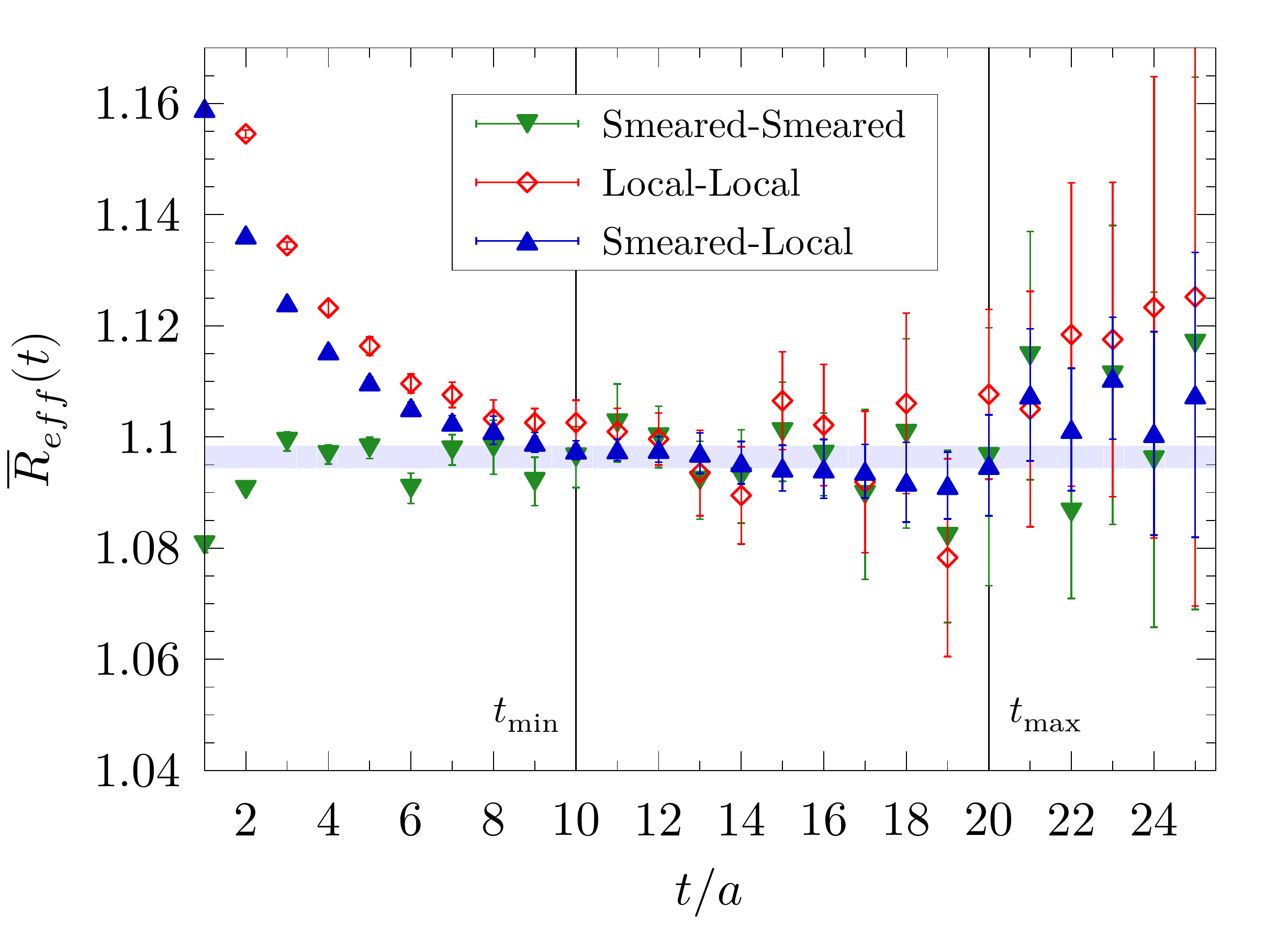}}}
\scalebox{1}{\subfigure[\it]{\includegraphics[width=8.15cm]{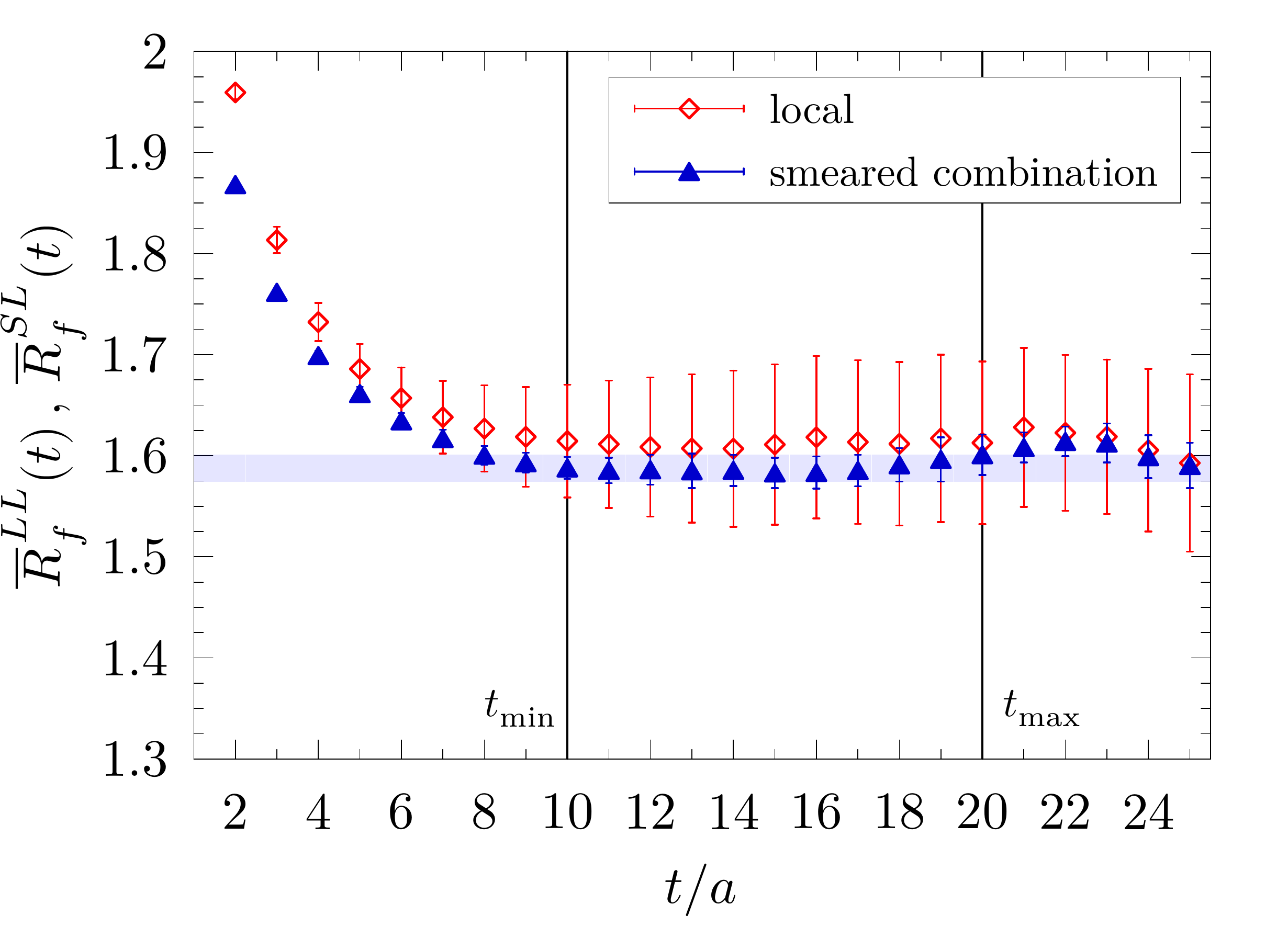}}}
\caption{\small \it (a) Effective mass ratio (\ref{Reff}) versus the Euclidean time distance $t$ corresponding to the $LL$, $SL$ and $SS$ correlators, calculated for $D_s$ and $D_s^*$ mesons in the case of the ETMC gauge ensemble A40.32. (b) The local (\ref{Rf_LL}) and smeared (\ref{Rf_SL}) ratios versus the Euclidean time distance $t$ for the same gauge ensemble. The horizontal bands represent the values of $R_s^M$ and $R_s^f$, including the statistical uncertainties at one standard deviation level, obtained from the plateau regions indicated by the vertical dotted lines.}
\label{fig:LLvsSL}
\end{figure}

In Fig.~\ref{fig:LLvsSL} we show the quality of the plateaux of the ratios (\ref{Reff}), (\ref{Rf_LL}) and  (\ref{Rf_SL}) in the case of the gauge ensemble A40.32 with $a\mu_\ell \simeq a\mu_s$ and $a\mu_h \simeq a\mu_c$. 
We also compare the extraction of masses and decay constants from Gaussian-smeared and/or local correlation functions. 
The smearing technique has two advantages: it allows the plateaux to be reached at earlier time distances and it improves the signal to noise ratio at large time distances, particularly in the case of the $SL$ correlation functions.
Thus, we have chosen the latter ones in order to extract the ground-state masses and decay constants in our analysis.

The $SS$ correlation functions exhibit the most anticipated plateaux.
Therefore, we make use of them to select the proper plateau range $t_{\mathrm{min}} \leq t \leq t_{\mathrm{max}}$, where the ground state can be considered safely isolated.
The value of $t_{\mathrm{min}}$ is identified as the point where the ratio $R^M_{\ell}$, obtained from $SL$ and $SS$ correlators, begin to intercept each other. 
On the other hand, the value of $t_{\mathrm{max}}$ is chosen in order to cut the largest statistical fluctuations.
It turns out that $t_{\mathrm{max}}$ gets smaller for heavier heavy-quark masses, although we checked that its choice has a negligible impact on the final results.
The plateaux ranges are given in Table~\ref{tab:tmin}. 
We have chosen common plateaux ranges for masses and decay constants, as they are extracted from the same correlators, and we don't observe any significant dependence on the light- or heavy-quark masses.
As a further check of the correct isolation of the ground-state, we have employed the GEVP method \cite{Blossier:2009kd}, which simultaneously involves the four correlators $LL, LS, SL$ and $SS$. 
The GEVP method yield results in agreement with those obtained using only the $SL$ correlators with a slightly larger uncertainty. 
 
\begin{table}[h!]
\begin{center}
\begin{tabular}{||c|c|c|c||}
\hline
$\beta$ & $L^3\times T$ & $t_{\mathrm{min}}/a$ & $t_{\mathrm{max}}/a$\\
\hline
\hline
1.90 & $32^3\times 64$ & 10 & 20\\
\cline{2-4}
     & $24^3\times48$ & 10 & 18\\
\hline
1.95 & $32^3\times 64$ & 12 & 20\\
\cline{2-4}
     & $24^3\times 48$ & 12 & 18\\
 \hline
2.10 & $48^3\times 64$ & 16 & 36\\
\hline
\end{tabular}
\end{center}
\caption{\it Values of $t_{\mathrm{min}}$ and $t_{\mathrm{max}}$ chosen to extract the ground-state signal from the $R_{\ell}^{M(f)}$ ratios constructed from the $SL$ and $SS$ correlators (see text).}
\label{tab:tmin} 
\end{table}

\section{$\mathbfit{D^*_{(s)}}$ mesons masses and decay constants}
\label{sec:FMD}
\begin{figure}[tb!]
\centering
\scalebox{1}{\includegraphics[width=8cm]{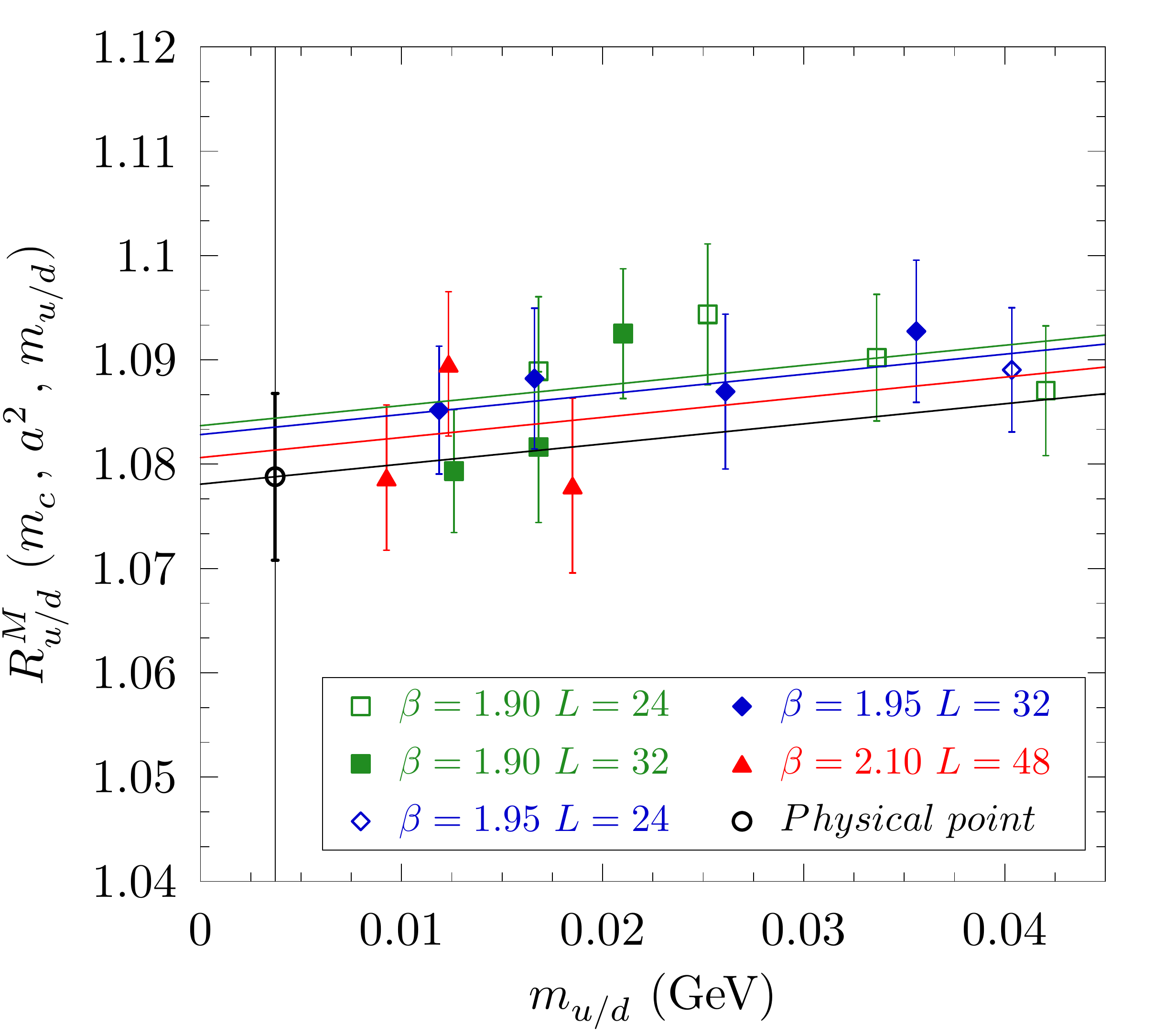}}
\scalebox{1}{\includegraphics[width=8cm]{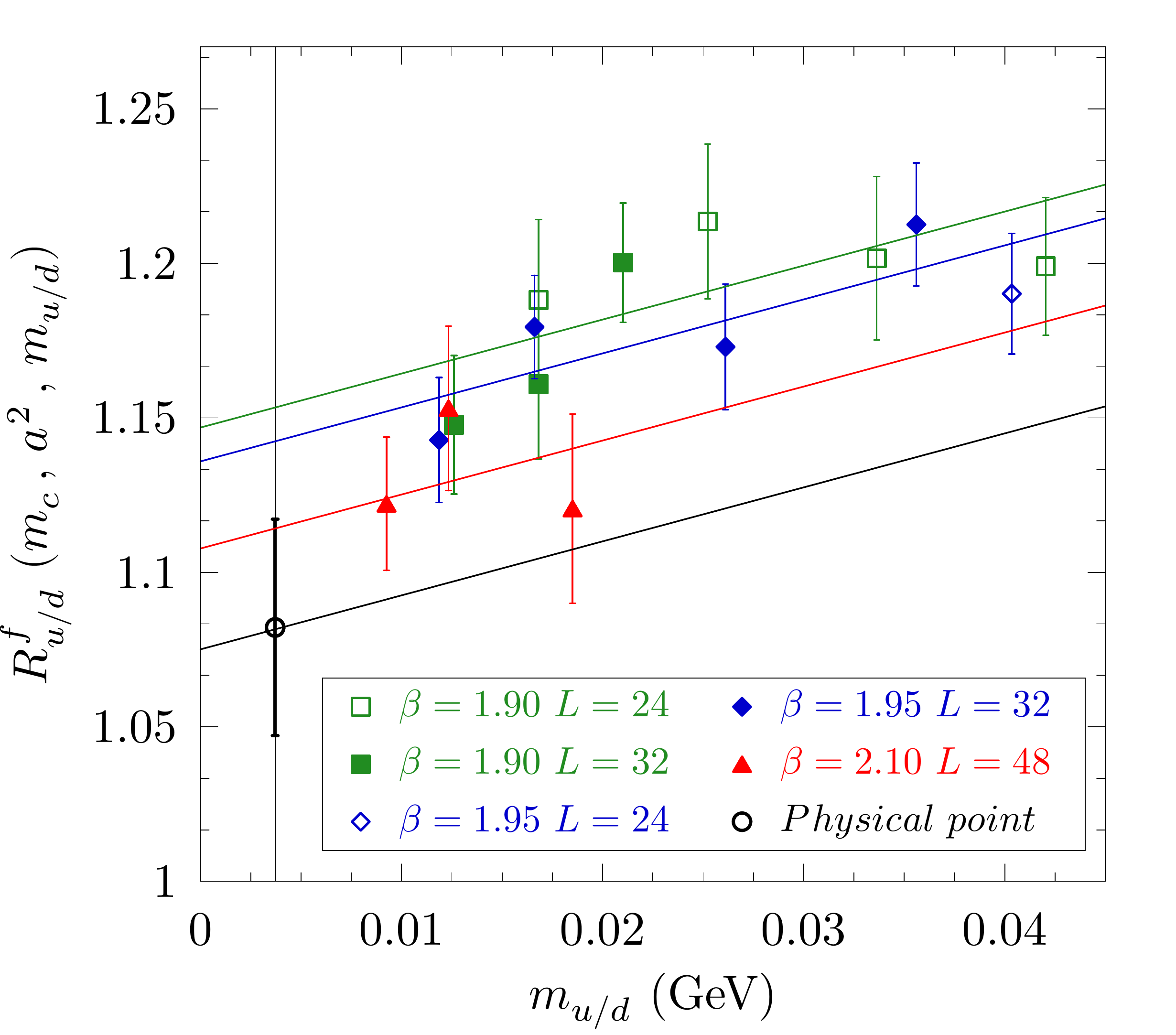}}
\scalebox{1}{\includegraphics[width=8cm]{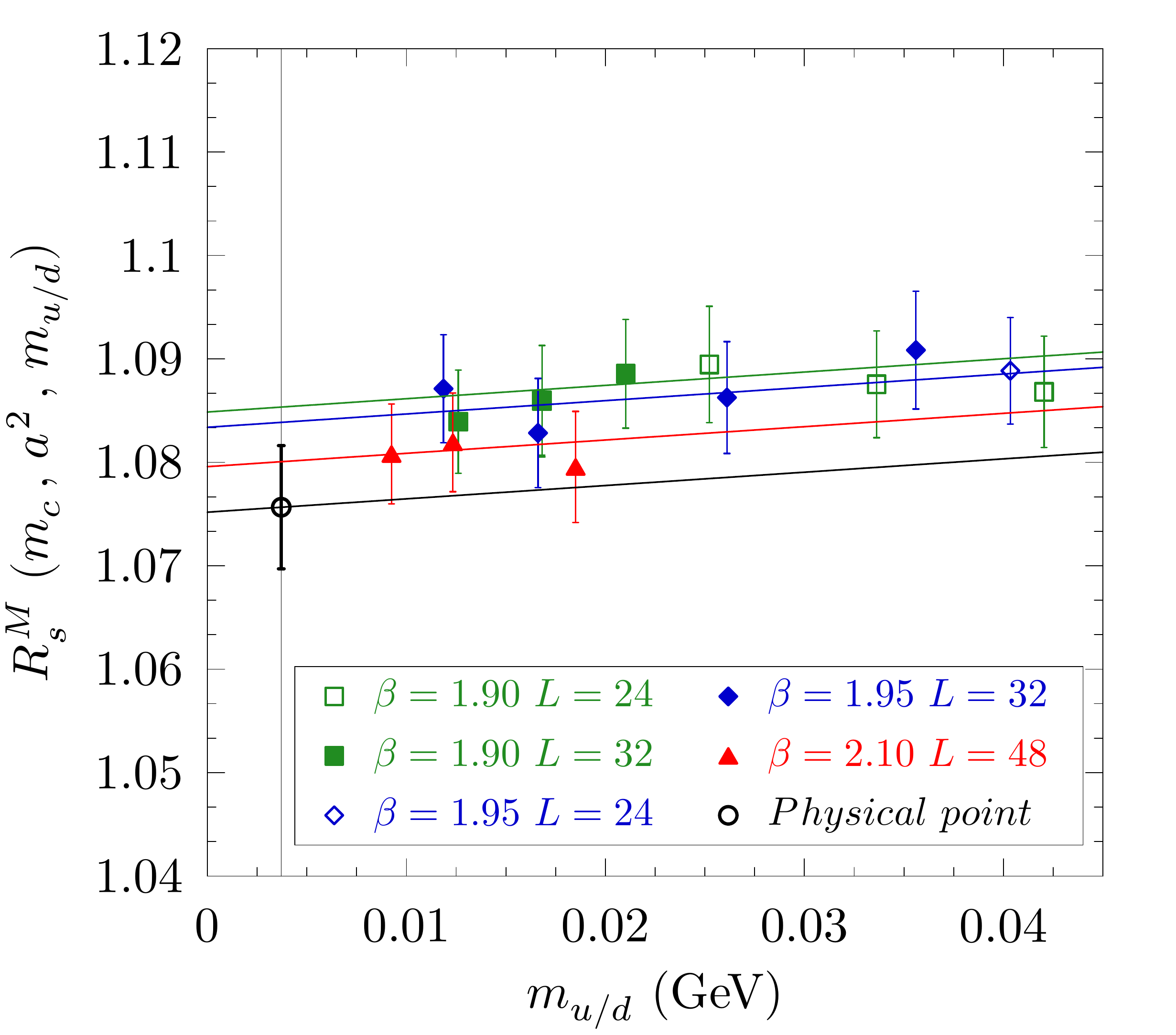}}
\scalebox{1}{\includegraphics[width=8cm]{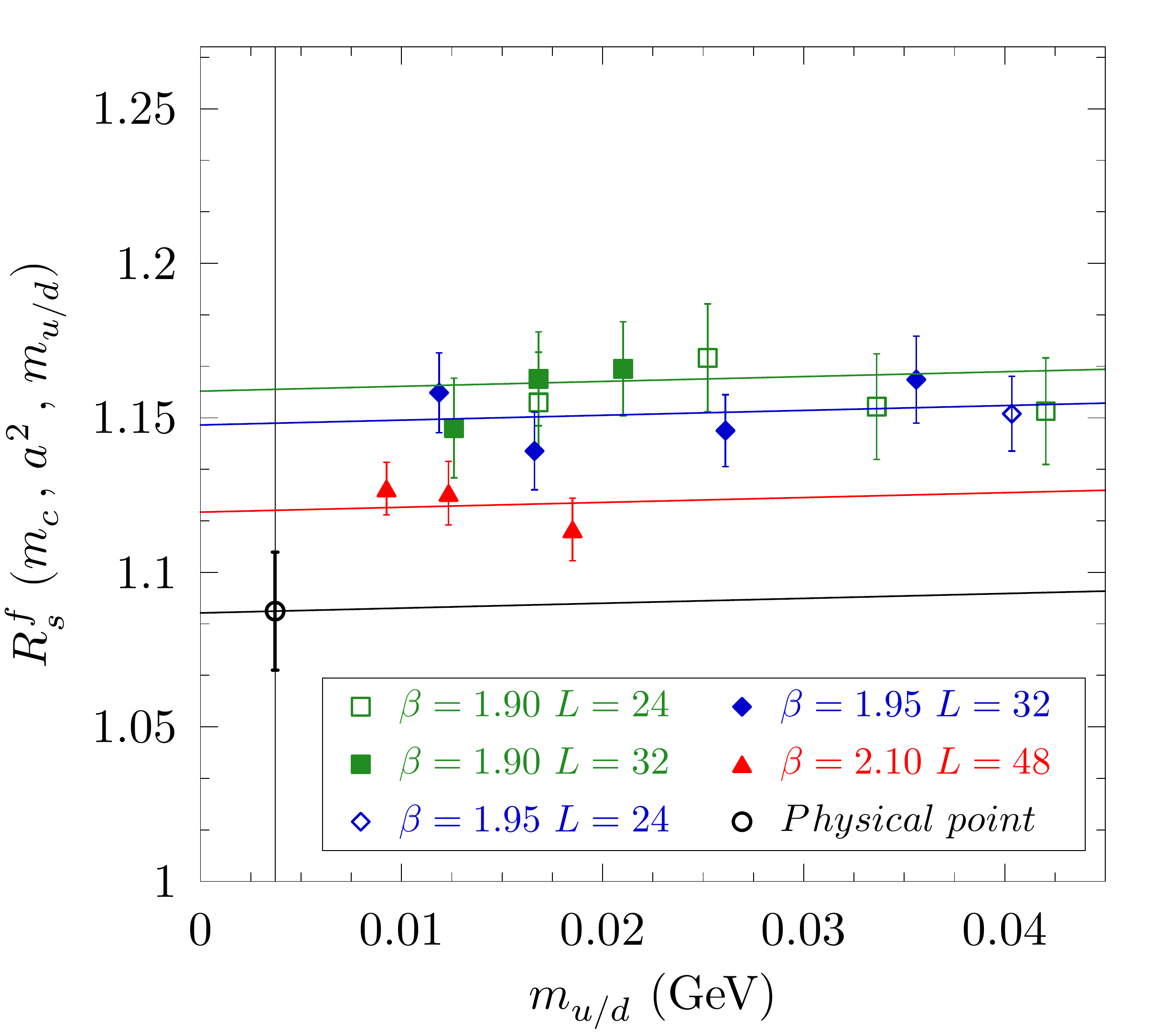}}
\caption{\small  \it Chiral and continuum extrapolations of the ratios $R^M_\ell(m_c)$ and $R^f_\ell(m_c)$ for $\ell = (u/d, s)$ based on the polynomial fit (\ref{fit}) with $P_3 = P_4 = 0$. The black points represent the values at the physical point $(m_{u/d}, a) = (m_{u/d}^{\mathrm{phys}}, 0)$.}
\label{fig:ContD} 
\end{figure}

We now present our analysis of the vector masses and decay constants in the charm sector.
We perform a smooth interpolation of the lattice data for the ratios $R_\ell^M$ and $R_\ell^f$ to the value of the physical charm quark mass $m_c^{\mathrm{phys}}(\overline{\mathrm{MS}},~2~\mbox{GeV}) = 1.176(39)$ GeV~\cite{Carrasco:2014cwa} and, for $\ell = s$, also to the value of the physical strange quark mass $m_s^{\mathrm{phys}}(\overline{\mathrm{MS}},~2~\mbox{GeV}) = 99.6(4.3)$ MeV~\cite{Carrasco:2014cwa}.
The dependencies of $R_\ell^M(m_c)$ and $R_{\ell}^f(m_c)$ on the renormalized up/down quark mass $m_{u/d} = a\mu_{u/d} / (aZ_P )$ and on the lattice spacing $a$ is investigated by performing a combined chiral and continuum extrapolation, based on a polynomial expansion of the form
 \be
    R^{fit}(a, m_{u/d}) = P_0 + P_1 m_{u/d} + P_2 a^2 + P_3 m_{u/d}^2 + P_4 a^4 ~ ,
    \label{fit}
 \ee
where we have taken into account that, for our Mtm setup, the automatic $O(a)$-improvement implies that discretization effects involve only even powers of the lattice spacing. 
The results obtained with the quadratic $m_{u/d}^2$ and the quartic $a^4$ terms have not been included in the final average (which therefore corresponds to $P_3  = P_4 = 0$), but they have been considered in order to estimate the uncertainty related to the chiral and continuum extrapolation, respectively. 
The combined extrapolations are shown in Fig.~\ref{fig:ContD}, where the physical point corresponds to $(m_{u/d}, a) = (m_{u/d}^{\mathrm{phys}}, 0)$ with $m_{u/d}^{\mathrm{phys}}(\overline{\mathrm{MS}},~2~\mbox{GeV}) = 3.70 (17)$ MeV~\cite{Carrasco:2014cwa}.
Note that: ~ i) the dependencies of both $R_\ell^M$ and $R_{\ell}^f$ on the light-quark mass is mild, and ~ ii) the discretization effects are of the order of $\sim 1 (5) \%$ in the case of the mass (decay constant) ratio.

In this way for the $D_{(s)}^{(*)}$ mesons we get
 \bea
    \label{MDstar}
    M_{D^*}/M_D & = & 1.0769\,(71)_{stat} (30)_{input}(13)_{tmin} (8)_{disc} (5)_{chir}\,[79] ~ , \\
    \label{MDstars}
    M_{D^*_s}/M_{D_s} & = & 1.0751(49)_{stat} (27)_{input} (4)_{tmin} (8)_{disc} (2)_{chir}\,[56] ~ , \\
    \label{fDstar}
    f_{D^*}/f_{D} & = & 1.078\,(31)_{stat} (5)_{input} (6)_{tmin} (8)_{disc} (9)_{chir} \,[36] ~ , \\
    \label{fDstars}
    f_{D^*_s}/f_{D_s} & = & 1.087\,(16)_{stat} (6)_{input} (6)_{tmin}  (7)_{disc} (5)_{chir}\,[20] ~ ,
 \eea
where the total uncertainty (in the square brakets) is the sum in quadrature of the statistical and various systematic uncertainties, which have been estimated in the following way: 
\begin{itemize}
\item the uncertainty labelled $tmin$ is computed by repeating the analysis with a value of $t_{\mathrm{min}}$ shifted by two units and taking half of the difference with the central values;                                                            
\item the chiral and discretization uncertainties, labelled respectively as $chir$ and $disc$, are obtained by considering either $P_3 \neq 0$ or $P_4 \neq 0$ in Eq.~(\ref{fit}) and taking again half of the difference with the central values; 
\item the uncertainty labelled $input$ comes from the uncertainties of the input parameters of Table~\ref{tab:boots}.
\end{itemize}
Combining our results (\ref{MDstar}-\ref{MDstars}) for $M_{D_{(s)}^*} /M_{D_{(s)}}$ with the experimental values of the $D_{(s)}$-meson masses~\cite{PDG} (used to calculate $m_c^{phys}$ in Ref.~\cite{Carrasco:2014cwa}) we obtain 
\be
     M_{D^*} = 2013 ~ (14) ~ \mbox{MeV} \hspace{5mm} \mathrm{and} \hspace{5mm} M_{D_s^*} = 2116 ~ (11) ~ \mbox{MeV} ~ ,
\ee
which compare well with the experimental values $M_{D^*}^{exp} = 2010.27 (5)$ MeV and $M_{D_s^*}^{exp} = 2112.1 (4)$ MeV \cite{PDG}. 

As for the decay constants, existing lattice calculations for $f_{D_{(s)}^*} /f_{D_{(s)}}$ have been carried out only with $N_f = 2 + 1$ and $N_f = 2$ dynamical quarks. 
The $N_f = 2 + 1$ estimate $f_{D_s^*} / f_{D_s} = 1.10 (2)$ \cite{DavisD} is in good agreement with our result, while the $N_f = 2$                                                                                                                                                                                                                                                                                                                                                                                                                                                                                                                                                                                                                                                                                                 results $f_{D^*} / f_D = 1.208 (27)$~\cite{sanfilippo_noi} and $f_{D_s^*} /f _{D_s} = 1.26( 3)$\cite{incriminato} are $\simeq 10\%$ larger than our predictions (\ref{fDstar}-\ref{fDstars}). 

Using the values of the pseudoscalar decay constants $f_D = 207.4 (3.8)$ MeV and $f_{D_s} = 247.2 (4.1)$ MeV determined by our collaboration in Ref.~\cite{leptonic} we get 
\be
     f_{D^*} = 223.5 ~ (8.7)~ \mbox{MeV} \hspace{5mm} \mathrm{and} \hspace{5mm} f_{D_s^*}= 268.8 ~ (6.5) ~ \mbox{MeV} ~ .
 \ee

\section{$\mathbfit{B^*_{(s)}}$ mesons masses and decay constants}
\label{sec:FMB}
\begin{figure}[tb!]
     \begin{minipage}[l]{6.5cm}
	   \includegraphics[width=6.5cm]{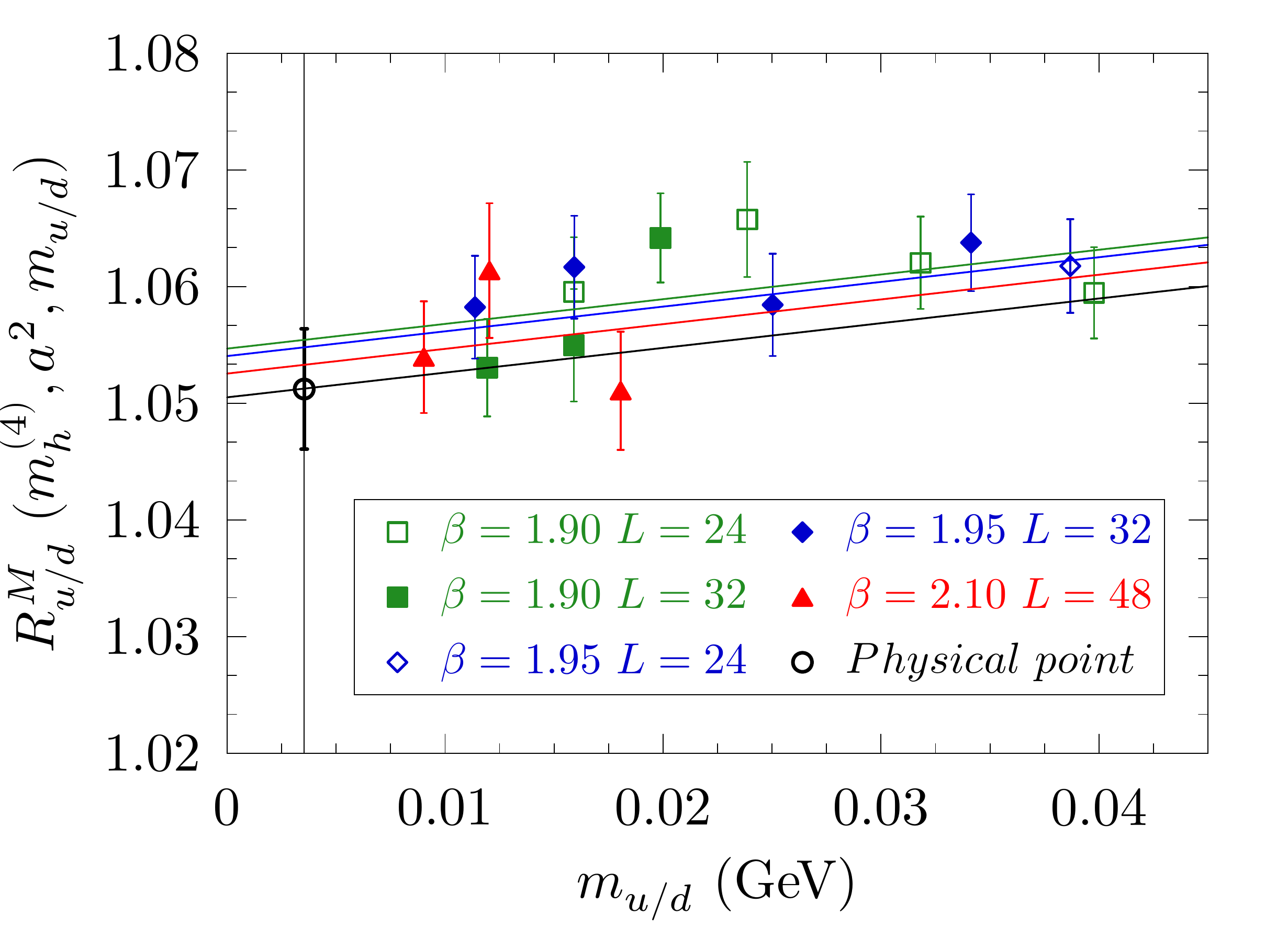}
	   \subfigure[\it]{\includegraphics[width=6.5cm]{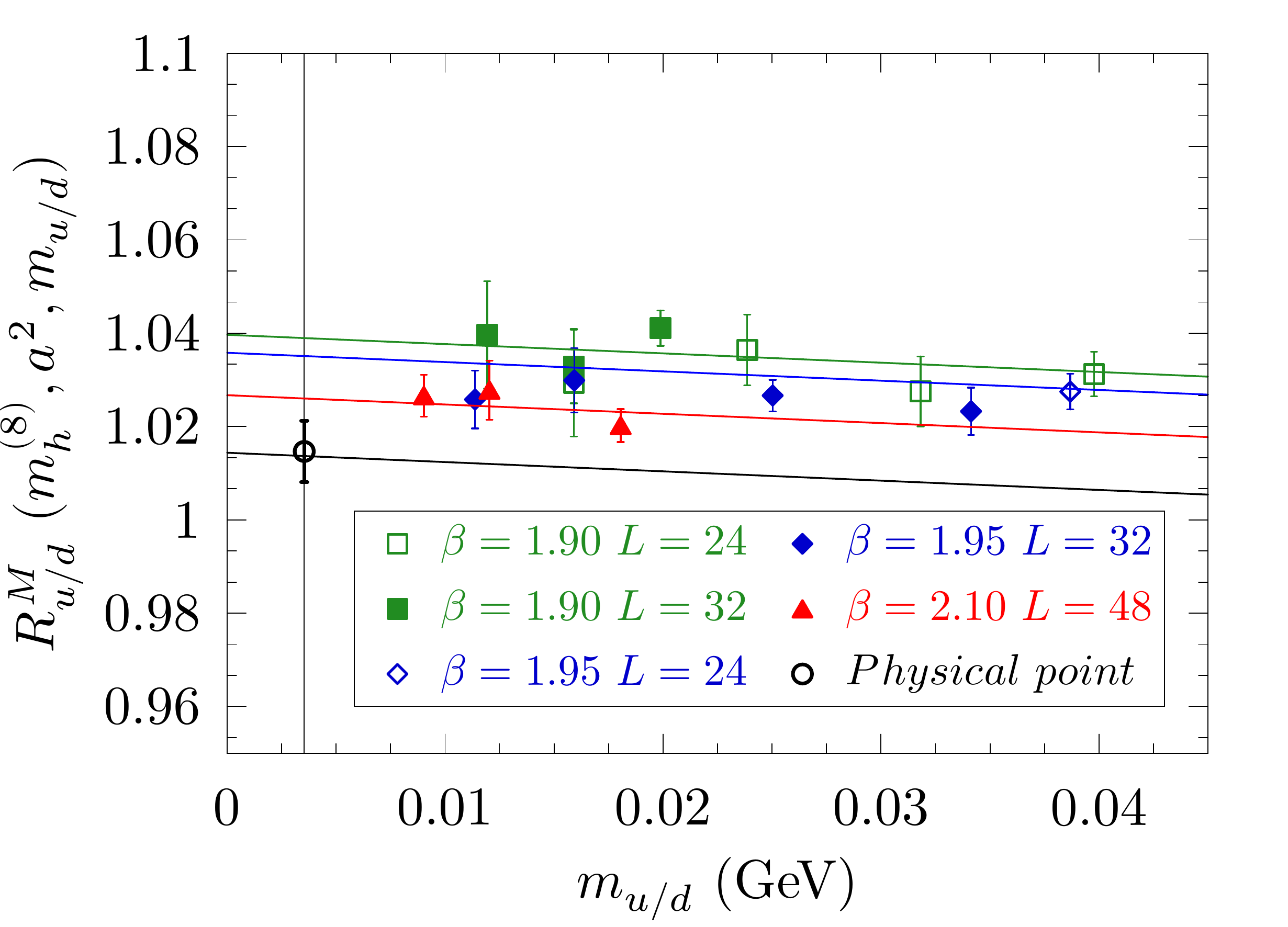}}
	\end{minipage}
	\begin{minipage}[r]{9.5cm}
	\vspace{1cm}
	 ~~ \subfigure[\it]{\includegraphics[width=9.5cm]{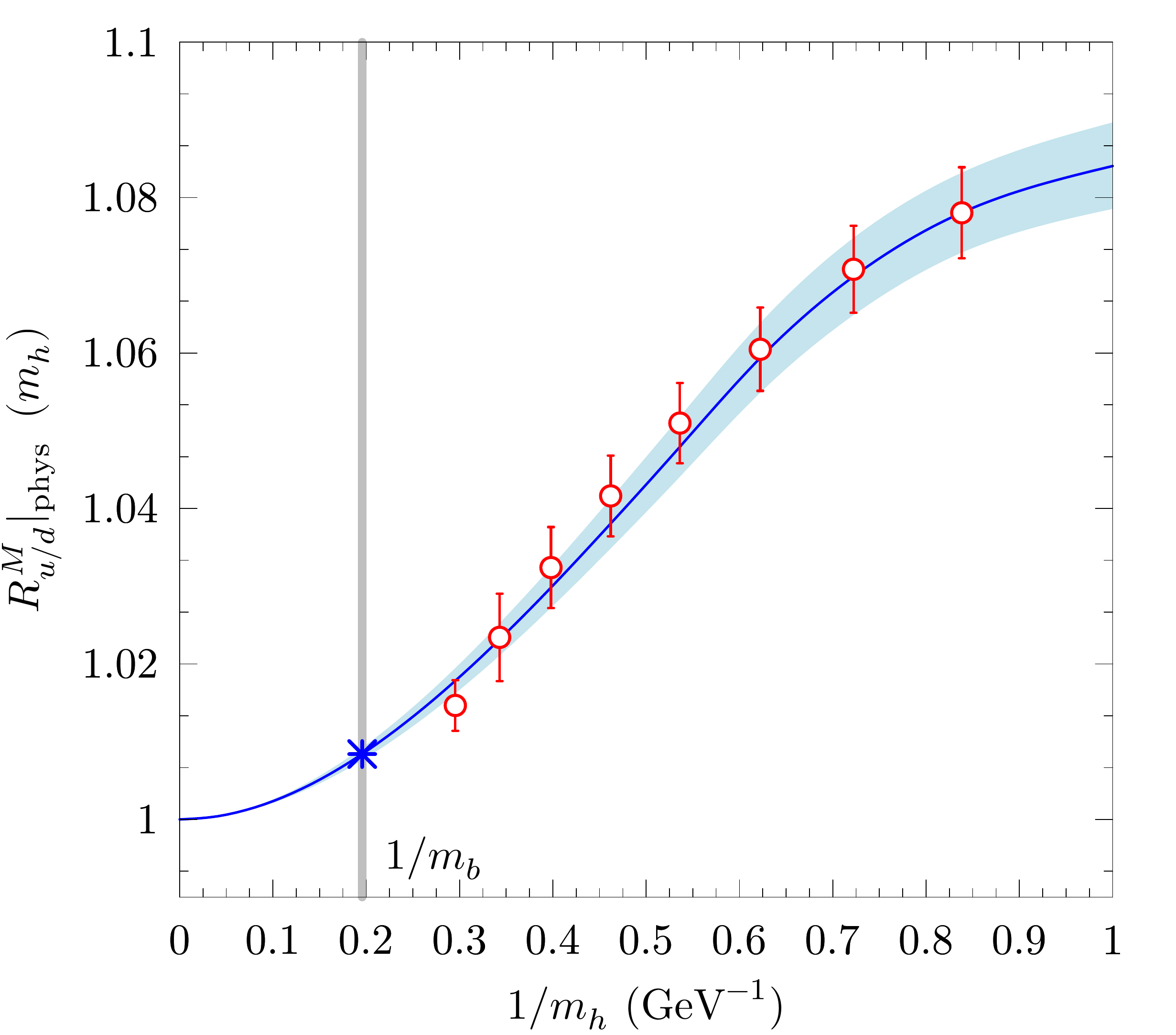}}
     \end{minipage}
\caption{\it (a) Chiral and continuum extrapolations of $R^M_{u/d}(m_h^{(k)})$ for $k = 4$ and $8$, based on the polynomial fit (\ref{fit}) with $P_3 = P_4 = 0$. (b) Dependence of $R_{u/d}^M |_{\rm phys}$ on the inverse heavy-quark mass $1 / m_h(\overline{MS},~2~\mbox{GeV})$ and its interpolated value at the physical $b$-quark mass. The interpolation is based on correlated fits according to Eq.~(\ref{fitMh}). The band corresponds to the fit uncertainty at one standard deviation. The vertical dotted line corresponds to $1 / m_b^{\mathrm{phys}}$ determined in Ref.~\cite{Bussone:2016iua}.}
\label{fig:M_B}
\end{figure}

\begin{figure}[tb!]
     \begin{minipage}[l]{6.5cm}
	   \includegraphics[width=6.5cm]{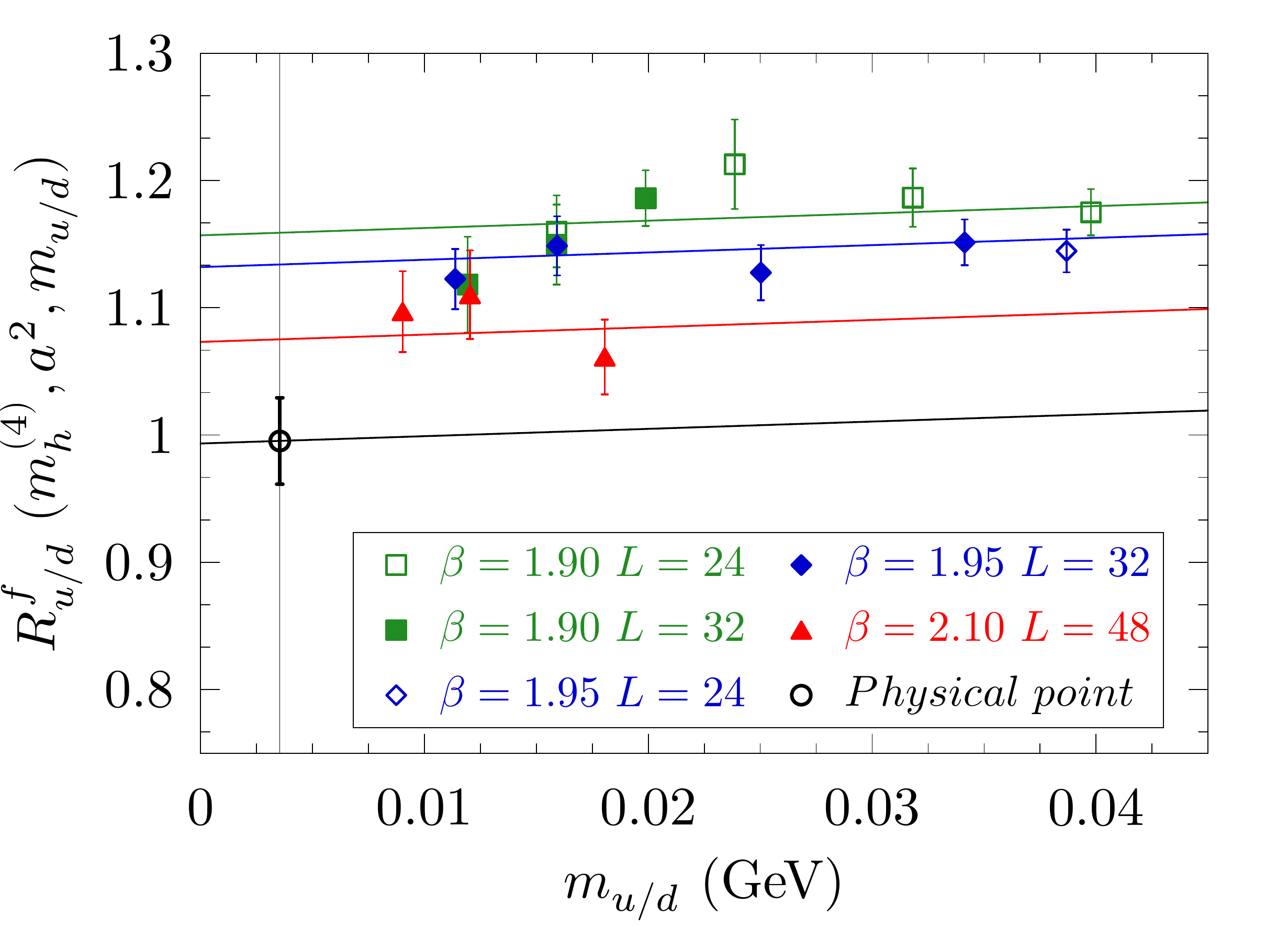}
	   \subfigure[\it]{\includegraphics[width=6.5cm]{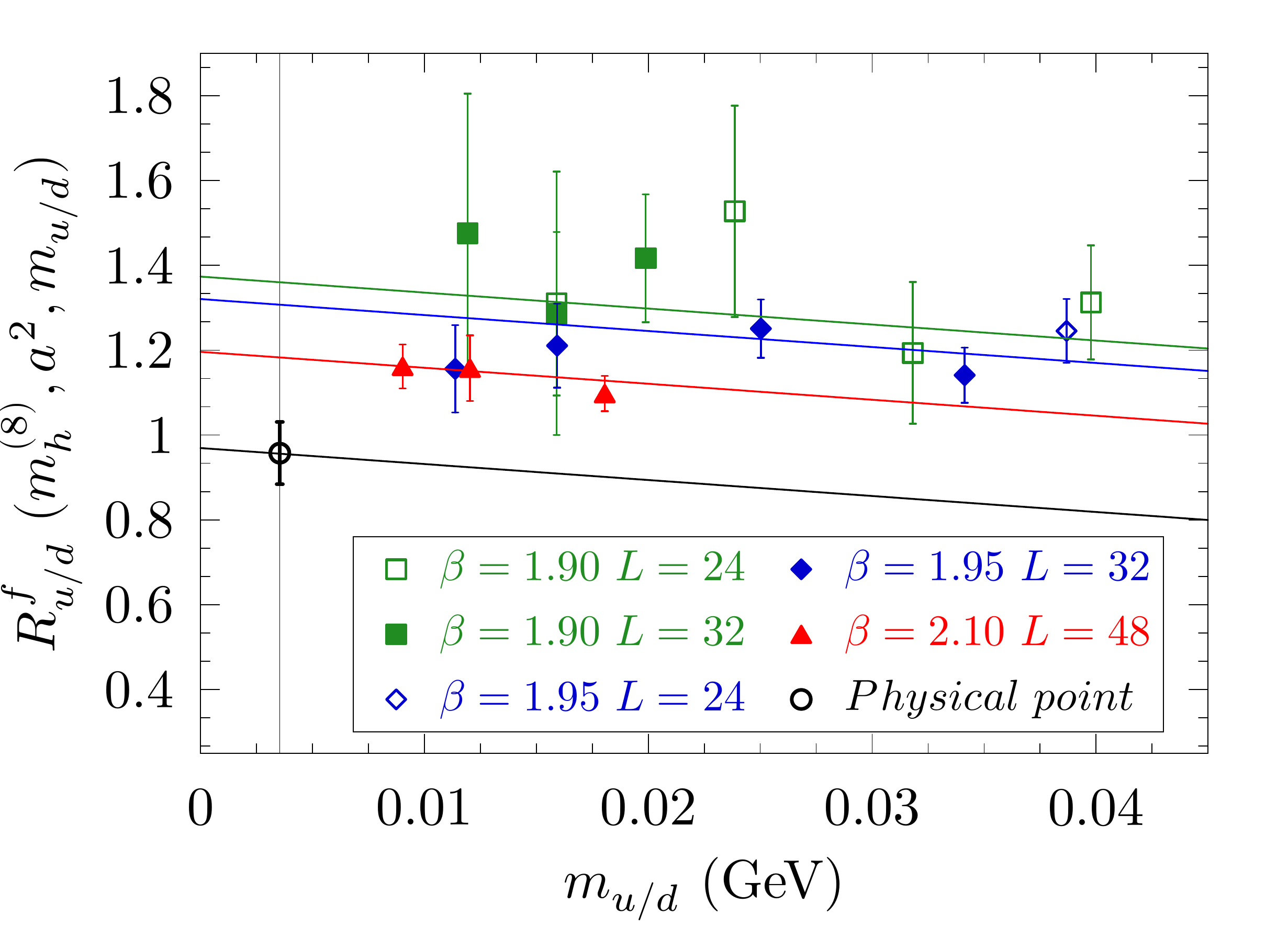}}
	\end{minipage}
	\begin{minipage}[r]{9.5cm}
	\vspace{1.1cm}
	 ~~ \subfigure[\it]{\includegraphics[width=9.5cm]{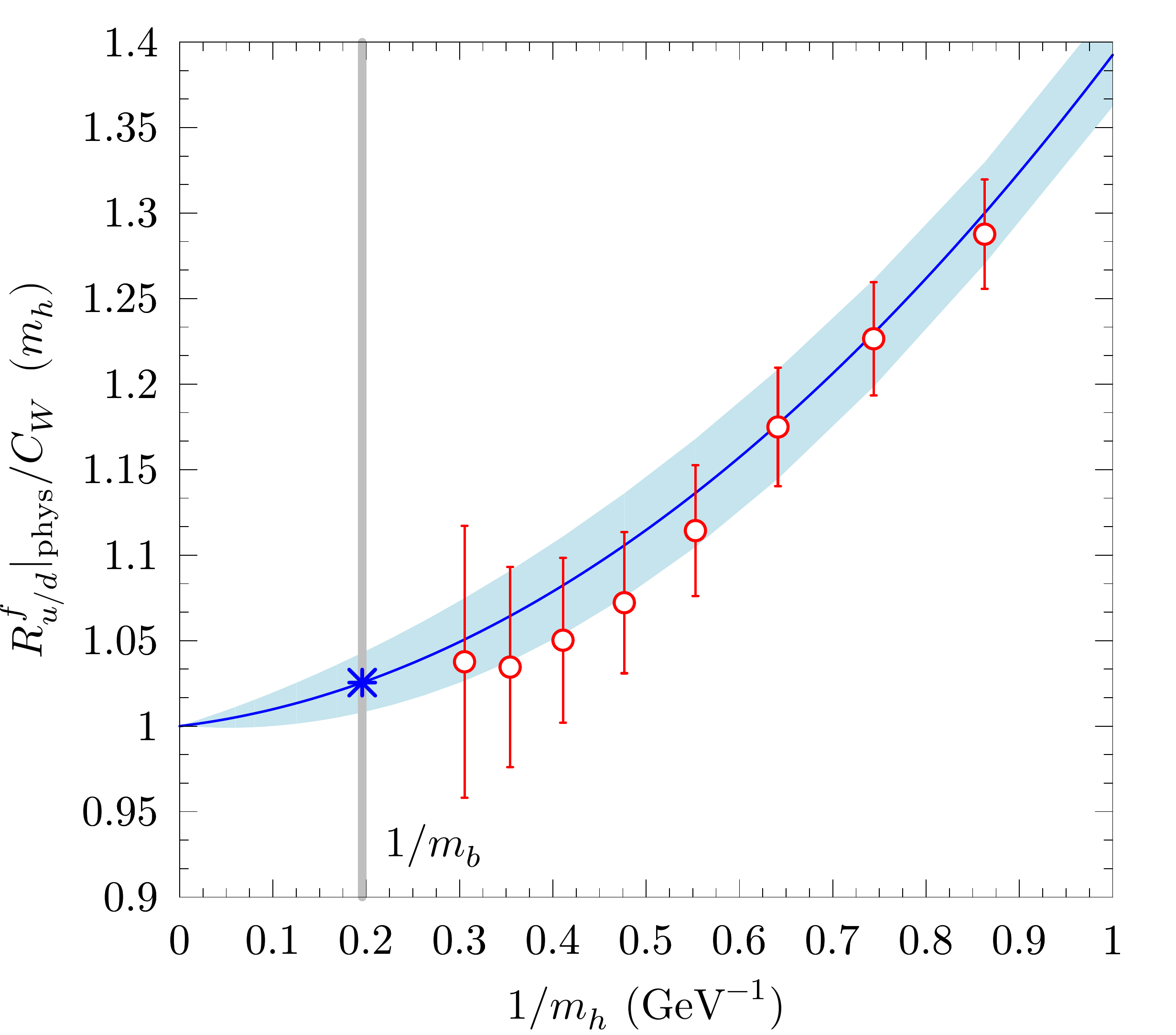}}
     \end{minipage}
\caption{\it Same as in Fig.~\ref{fig:M_B}, but for the ratios $R_{u/d}^f(m_h^{(k)})$ (a) and $R_{u/d}^f |_{\rm phys} / C_W(m_h)$ (b).}
\label{fig:f_B}
\end{figure}

\begin{figure}[tb!]
     \begin{minipage}[l]{6.5cm}
	   \includegraphics[width=6.5cm]{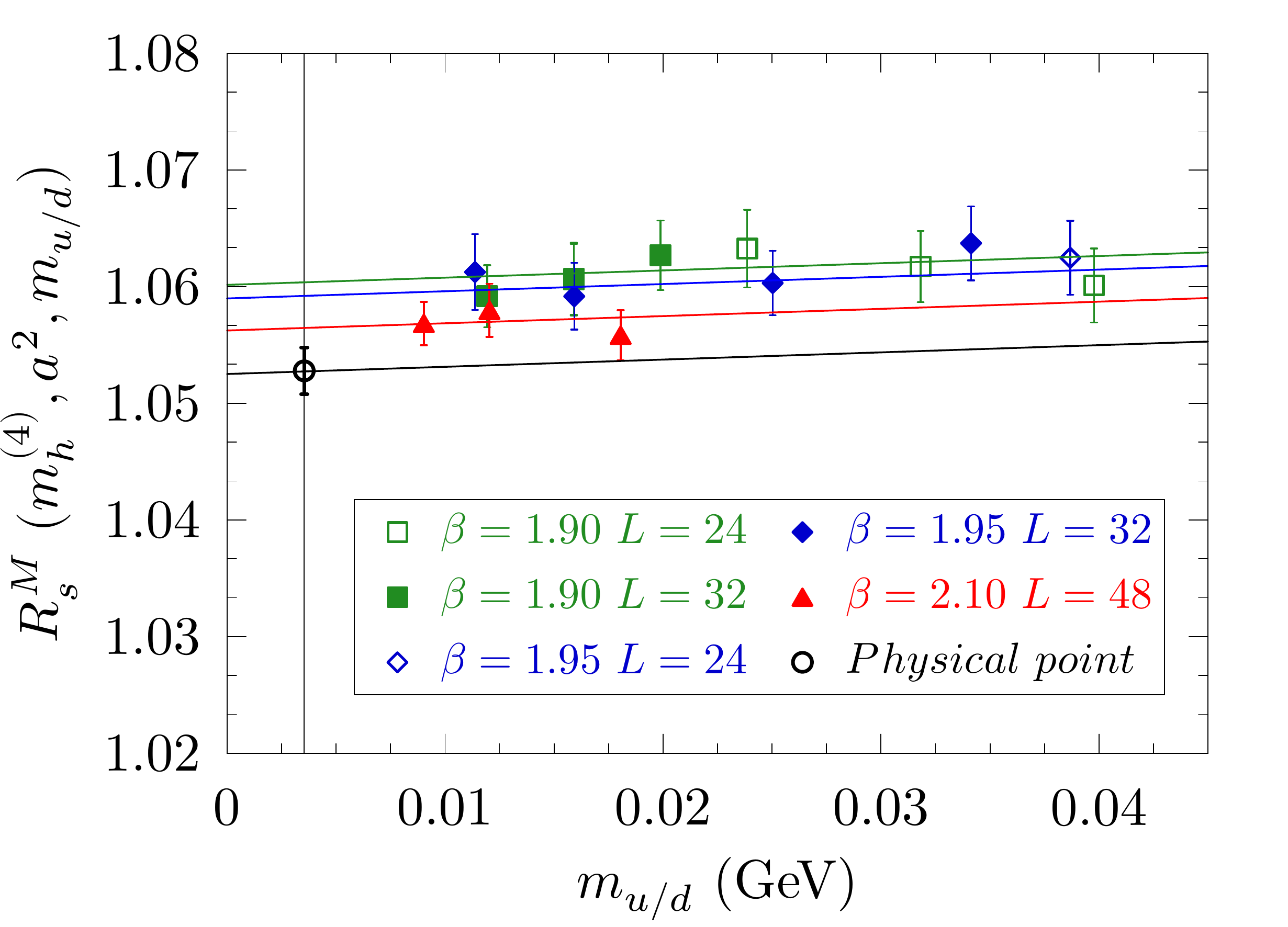}
	   \subfigure[\it]{\includegraphics[width=6.5cm]{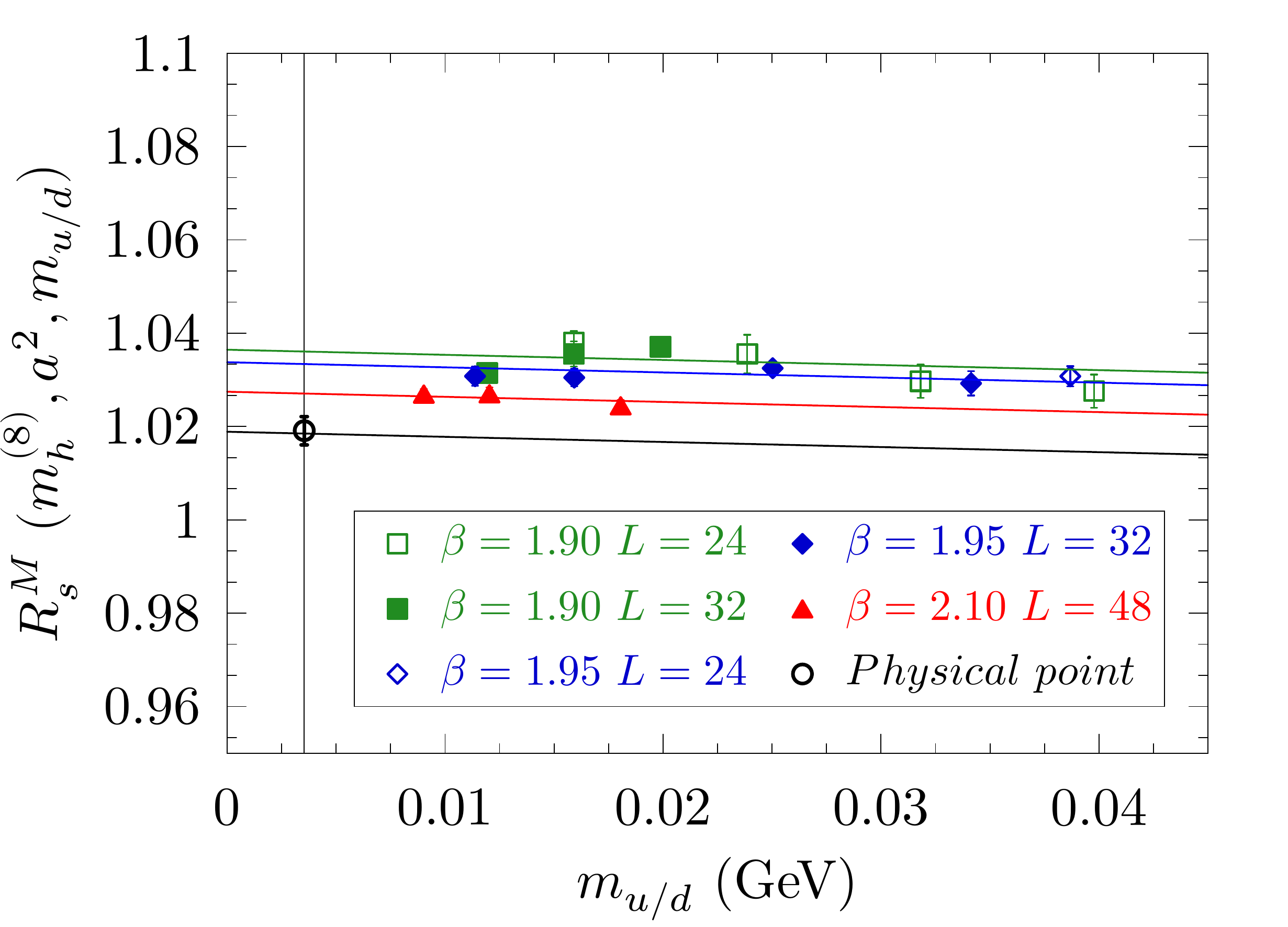}}
	\end{minipage}
	\begin{minipage}[r]{9.5cm}
	\vspace{1.1cm}
	 ~~ \subfigure[\it]{\includegraphics[width=9.5cm]{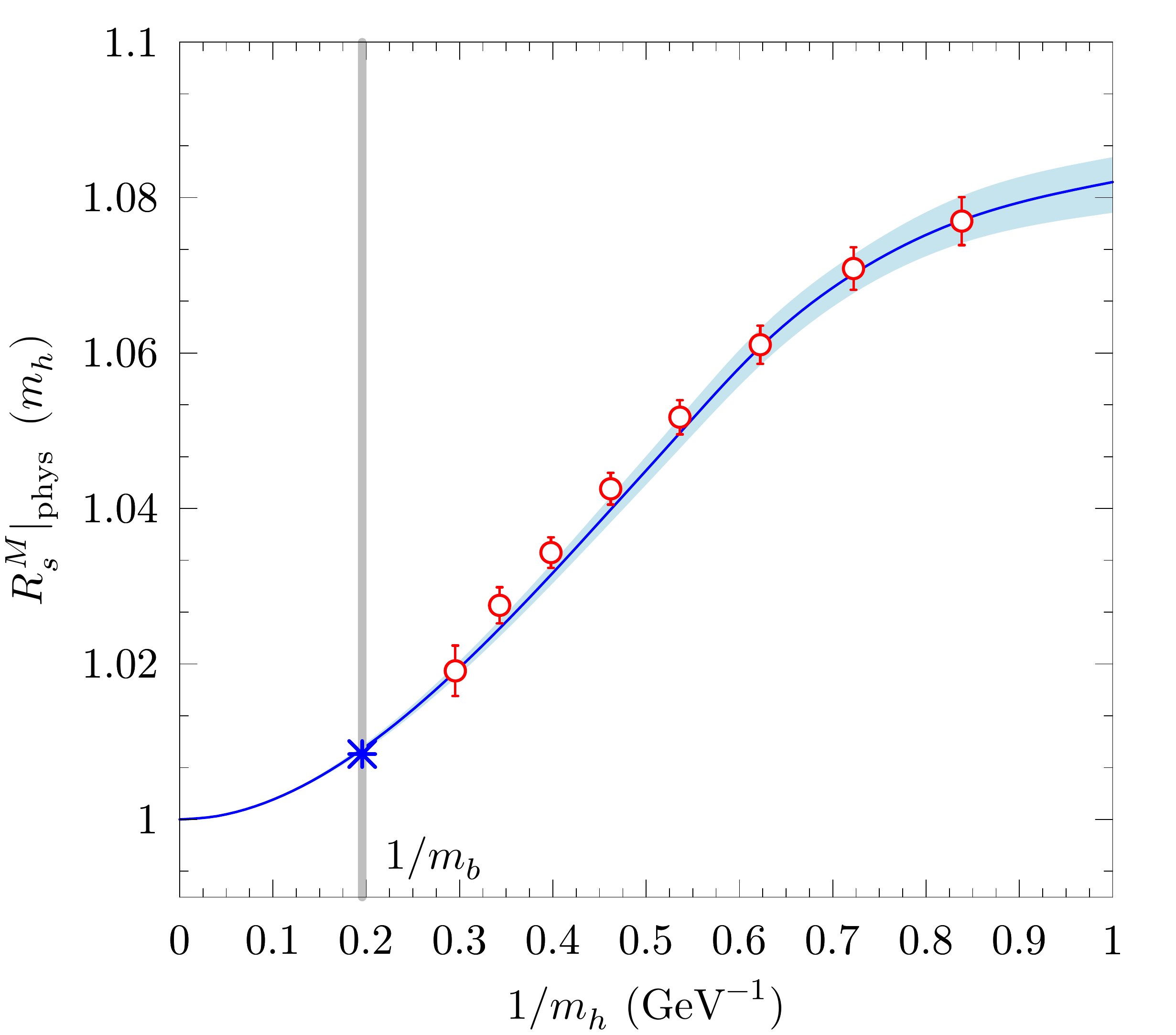}}
     \end{minipage}
     \vspace{-0.3cm}\caption{\it Same as in Fig.~\ref{fig:M_B}, but for the ratios $R_s^M(m_h^{(k)})$ (a) and $R_s^M |_{\rm phys}(m_h)$ (b).}
\label{fig:M_Bs}
\vspace{0.5cm}
     \begin{minipage}[l]{6.5cm}
	   \includegraphics[width=6.5cm]{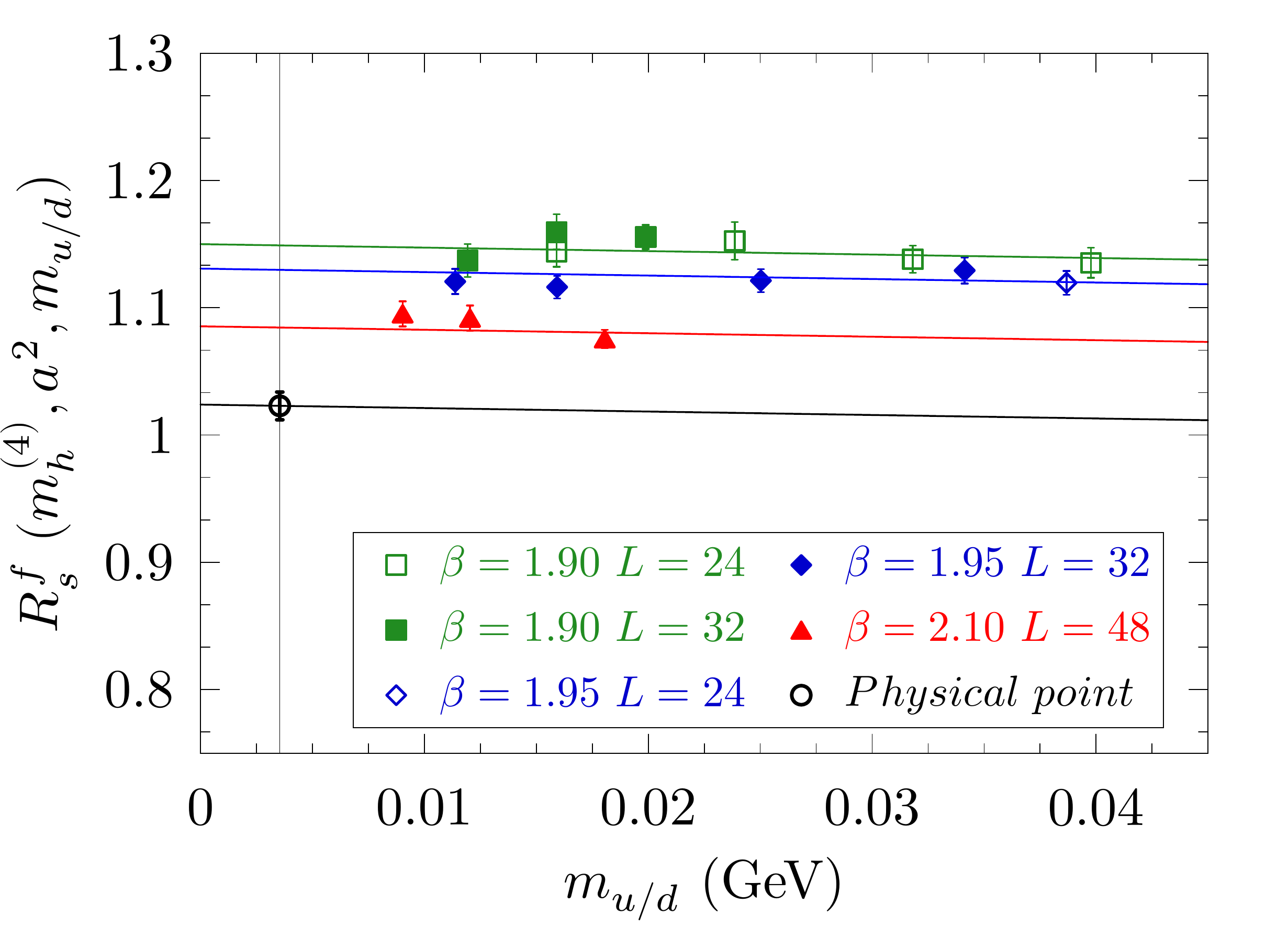}
	   \subfigure[\it]{\includegraphics[width=6.5cm]{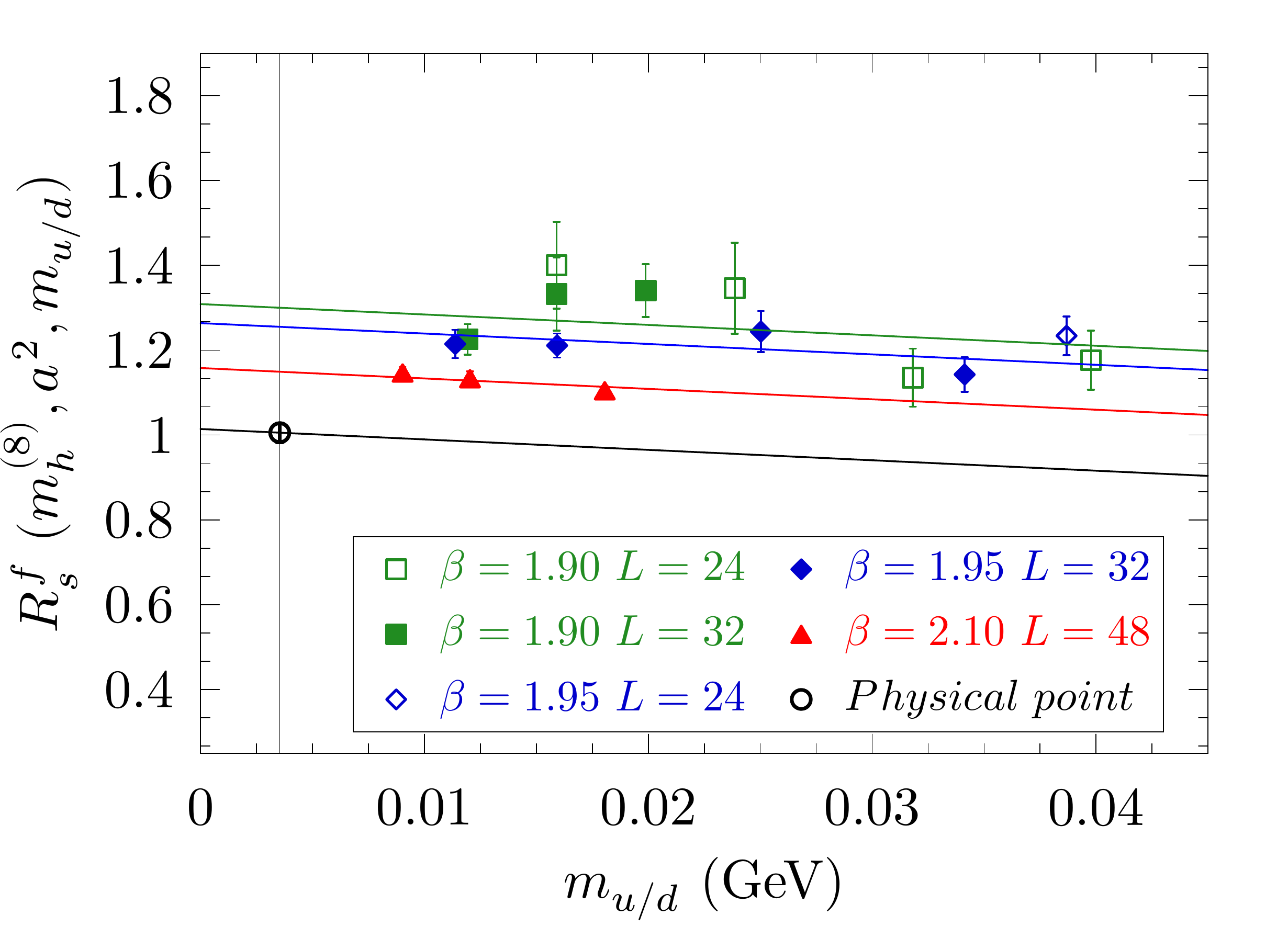}}
	\end{minipage}
	\begin{minipage}[r]{9.5cm}
	\vspace{1.1cm}
	~~ \subfigure[\it]{\includegraphics[width=9.5cm]{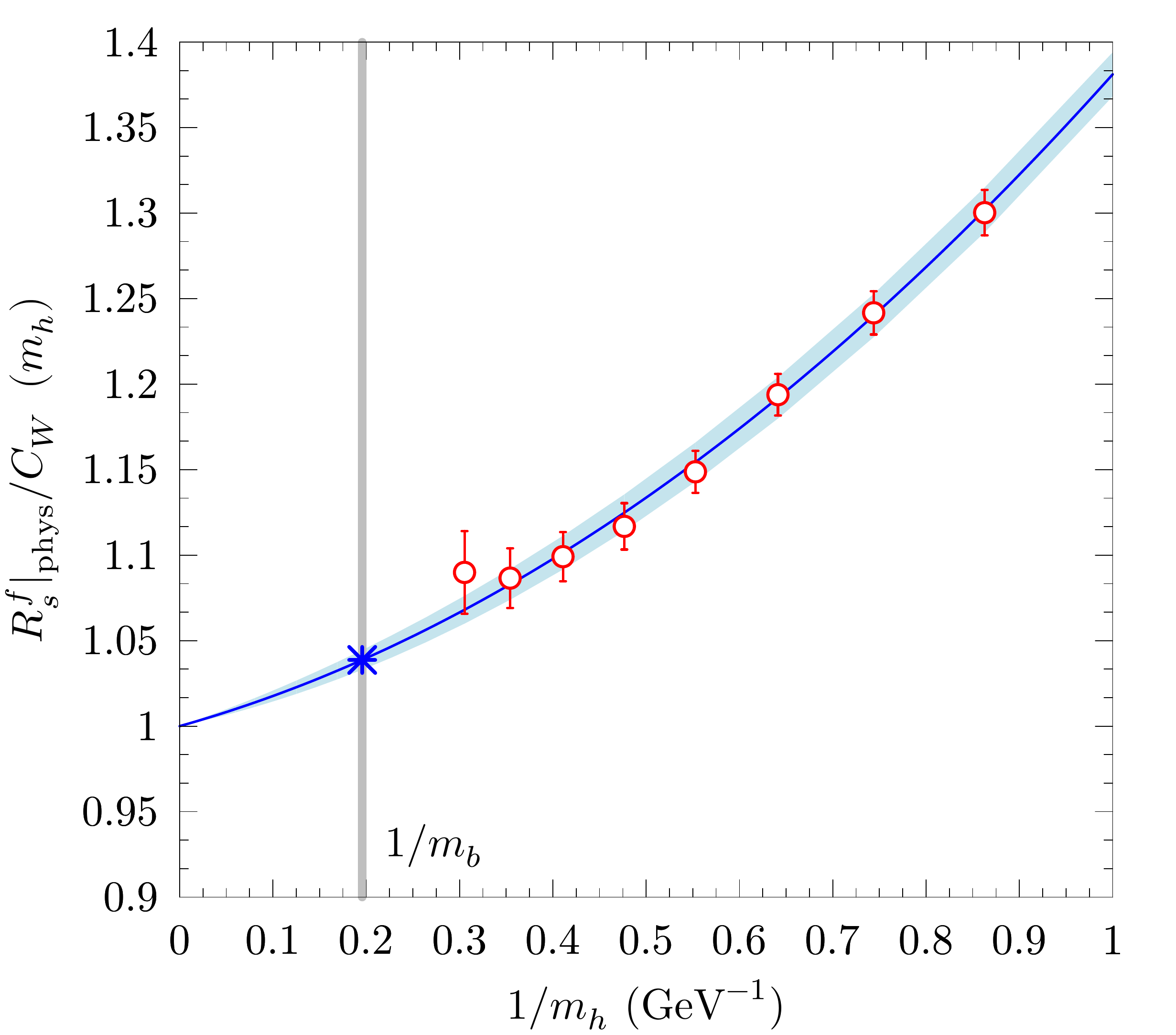}}
     \end{minipage}
     \vspace{-0.3cm}\caption{\it Same as in Fig.~\ref{fig:M_B}, but for the ratios $R_s^f(m_h^{(k)})$ (a) and $R_s^f |_{\rm phys} / C_W(m_h)$ (b).}
\label{fig:f_Bs}
\end{figure}

We now present our analysis of the vector masses and decay constats in the beauty sector.
We have computed the ratios $R_\ell^M(m_h)$ and $R_\ell^f(m_h)$ for a series of heavy quark masses $\{m_h^{(k)}\} \geq m_c$ with $k = 1, ..., 8$ (see Table~\ref{tab:param}). 
For each of these ratios the results are extrapolated to the chiral and continuum limits, as shown in the panels (a) of Figs.~\ref{fig:M_B}-\ref{fig:f_Bs} for some illustrative cases.
Note that: ~ i) the dependencies of both $R_\ell^M$ and $R_{\ell}^f$ on the light-quark mass is mild, and ~ ii) the discretization effects are of the order of $\sim 1 (5) \%$ in the case of the mass (decay constant) ratio.
This is similar to what has been observed for the charmed mesons (see Fig.~\ref{fig:ContD}) though the heavy-quark mass is higher than the charm mass.
In what follows we will indicate by $R_\ell^M |_{\rm phys}$ and $R_\ell^f |_{\rm phys}$ the extrapolated values of the corresponding ratios at the physical point $(m_{u/d}, a) = (m_{u/d}^{\mathrm{phys}}, 0)$.

The HQET predicts that the ratios $R_\ell^M |_{\rm phys}(m_h)$ for $\ell = (u/d, s)$ are equal to one in the static heavy-quark limit, i.e.
 \be
     \lim_{m_h \rightarrow \infty} R_\ell^M |_{\rm phys}(m_h) = 1  ~ .
     \label{constraint_M}
 \ee
For the ratios of decay constants the perturbative matching between QCD and HQET has to be taken into account.
Introducing the HQET ratios 
 \be
     \overline{R}_\ell^f |_{\rm phys}(m_h) \equiv \frac{R_\ell^f |_{\rm phys}(m_h)}{C_W(m_h)} ~ ,
     \label{eq:Rf_HQET}
 \ee
where $C_W(m_h)$ is given at next-to-next-leading order in the strong coupling constant by~\cite{matching}
\be
     C_W(m_h) = 1 - \frac{2}{3}\frac{\alpha_s(m_h)}{\pi} - \left[ -\frac{1}{9}\zeta(3) + \frac{2}{27}\pi^2\log2 + \frac{4}{81}\pi^2+\frac{115}{36} \right]
                           \left(\frac{\alpha_s(m_h)}{\pi}\right)^2 ~ ,
     \label{Cw}
 \ee
the HQET predicts that the ratios $\overline{R}_\ell^f |_{\rm phys}(m_h)$ for $\ell = (u/d, s)$ are equal to one in the static heavy-quark limit, i.e.
 \be
     \lim_{m_h \rightarrow \infty} \overline{R}_\ell^f |_{\rm phys}(m_h) = 1 ~ .
     \label{constraint_f}
 \ee

We then perform correlated polynomial fits in the inverse powers of the heavy-quark mass $m_h$ by imposing the static limit constraints (\ref{constraint_M}) and (\ref{constraint_f}), namely
 \bea
      \label{fitMh} 
      R_\ell^M |_{phys}^{fit} & = & 1 + D_2 / m_h^2 + D_3 / m_h^3  + D_4 / m_h^4 ~ , \\[2mm]
      \label{fitfh} 
      \overline{R}_\ell^f |_{phys}^{fit} & = & 1 + \overline{D}_1 / m_h + \overline{D}_2 / m_h^2 + \overline{D}_3 / m_h^3 ~ ,  
 \eea
where we have taken into account that, according to the HQET, the linear term is absent in the case of the mass ratios (i.e., $D_1 = 0$ in Eq.~(\ref{fitMh})).
The interpolations of the various ratios in the inverse heavy-quark mass are shown in Fig.~\ref{fig:M_B} together with the results corresponding to the physical $b$-quark mass $m_b^{\mathrm{phys}}\mathrm{(\overline{MS},2 GeV)} = 5.201 (90)$~\cite{Bussone:2016iua}.
Note that the fit uncertainty, shown as a band in Fig.~\ref{fig:M_B}(b), is close to the uncertainty of the data around the charm region, it decreases as the heavy-quark mass increases and it vanishes in the static limit.
This is due to the fact that HQET constraint (\ref{constraint_M}) has no error being exactly known (up to higher-order radiative corrections neglected in Eq.~(\ref{Cw})).

Our final results for the $B_{(s)}^{(*)}$ mesons are
 \bea
    \label{MBstar}
    M_{B^*}/M_B & = &1.0078\,(8)_{stat}(8)_{chir} (7)_{tmin}(5)_{disc}(2)_{input} ~ [14] ~ , \\
    \label{MBstars}
    M_{B^*_s}/M_{B_s} & = & 1.0083\,(6)_{stat}(7)_{chir}(6)_{disc}(3)_{tmin}(2)_{input} ~ [11] ~ , \\
    \label{fBstar}
    f_{B^*}/f_B & = & 0.958\,(18)_{stat} (10)_{disc}(6)_{chir}(5)_{tmin}(2)_{input} ~ [22] ~ , \\
    \label{fBstars}
    f_{B^*_{s}}/f_{B_{s}} & = & 0.974\,(7)_{stat}(6)_{disc}(3)_{tmin}(2)_{input} (1)_{chir} ~ [10] ~ ,
\eea
where the error budget accounts for the same sources of uncertainties already considered for the charm sector in Sec.~\ref{sec:FMD}. 
In addition, we have verified that the inclusion of higher order terms in $1/m_h$ in the fitting Ans\"atze (\ref{fitMh}-\ref{fitfh}) used for the heavy-quark interpolation has a negligible effect when compared to the other sources of uncertainty taken already into account.

Our results (\ref{MBstar}-\ref{MBstars}) for the $M_{B_{(s)}^*} / M_{B_{(s)}}$ mass ratios can be combined with the experimental values of $B_{(s)}$-meson masses \cite{PDG} (used to evaluate $m_b^{phys}$ in Ref.~\cite{Bussone:2016iua}) to obtain
 \be
      M_{B^*} = 5320.5 ~ ( 7.6) ~ \mbox{MeV} \hspace{5mm} \mathrm{and} \hspace{5mm} M_{B_s^*} = 5411.8 ~ (6.2) ~ \mbox{MeV} ~ ,
 \ee
which compare well with the experimental values $M_{B^*}^{exp} = 5324.83 (32)$ MeV and $M_{B_s^*}^{exp} = 5415.4 (1.6)$ MeV \cite{PDG}. 

As for the decay constant ratios, we can compare our results with a recent determination obtained by the HPQCD collaboration \cite{DavisB} with $N_f = 2 + 1 + 1$ dynamical quarks: $f_{B^*} / f_B = 0.941 (26)$ and $f_{B_s^*} / f_{B_s} = 0.953 (23)$. 
They can also be compared with a recent calculation based on the QCD sum rule approach of Ref.~\cite{Lucha:2015xua}, yielding $f_{B^*} / f_B = 0.944 (23)$ and $f_{B_s^*} / f_{B_s} = 0.947 (30)$. 
All the above estimates are nicely consistent with our results. 
On the contrary, as for the $D^*_{(s)}$ case, we still find a $\simeq 10\%$ difference with the $N_f = 2$ determination $f_{B^*}/f_B = 1.051(17)$ from Ref.~\cite{sanfilippo_noi}. 

Combining our results (\ref{fBstar}-\ref{fBstars}) with the pseudoscalar decay constants $f_B = 193 (6)$ MeV and $f_{B_s} = 229 (5)$ MeV, calculated by ETMC in Ref.~\cite{Bussone:2016iua}, yields the following predictions for the vector meson decay constants
 \be
      f_{B^*} = 186.4 ~ (7.1) ~ \mbox{MeV} \hspace{5mm} \mathrm{and} \hspace{5mm} f_{B_s^*} = 223.1 ~ (5.6) ~ \mbox{MeV} ~ .
 \ee

\section{Conclusions}
\label{sec:concl}
We have computed the masses and the decay constants of vector heavy-light mesons using the ETMC gauge configurations with $N_f = 2 + 1 +1$ dynamical quarks. 
Our results reproduce very well the experimental values of both $D_{(s)}^*$- and $B_{(s)}^*$-meson masses. 

We have found that $f_{D_{(s)}^*} / f_{D_{(s)}} > 1$ and $f_{B_{(s)}^*} / f_{B_{(s)}} < 1$ with a spin-flavor symmetry breaking effect of $\simeq +8 \%$ in the charm sector and $\simeq - 4 \%$ in the beauty sector. 
Our results for the decay constant ratio exhibit a tension with the corresponding lattice determinations obtained by ETMC at $N_f = 2$ \cite{incriminato,sanfilippo_noi}, while they are consistent with the findings of Refs.~\cite{DavisD,DavisB} obtained by HPQCD with $N_f = 2 + 1 (+1)$ dynamical quarks. 

Since our analysis follows almost the same steps of the ETMC analyses at $N_f = 2$ of Refs.~\cite{incriminato,sanfilippo_noi}, the observed $\simeq 10\%$ tension might be due to a dependence on the number of sea quarks, and in particular to the inclusion of the strange quark. 
The possibility that the observed difference can be attributed to a quenching effect of the strange quark is a quite interesting issue, because its size would be larger than what typically expected. 
Further investigations at different $N_f$ values are required in order to assess this issue.  

\section*{Acknowledgments}
We gratefully acknowledge the CPU time provided by PRACE under the project PRA067 {\it ``First Lattice QCD study of B-physics with four flavors of dynamical quarks"} on the BG/Q systems Juqueen at JSC (Germany) and Fermi at CINECA (Italy) and by CINECA under the specific initiative INFN-LQCD123. We thank MIUR (Italy) for partial support under the contract PRIN 2010-2011 and PRIN 2015.

\end{document}